\RequirePackage{ifpdf}

\documentclass{PoS}

\usepackage[authoryear,square]{natbib}
\bibpunct{(}{)}{;}{a}{}{,}
\usepackage{graphicx}
\usepackage{amsmath}
\usepackage{etoolbox}

\title{Cosmology with a SKA HI intensity mapping survey}

\ShortTitle{Cosmology with SKA HI IM surveys}

\author{\speaker{Mario G. Santos},$^{a,b}$ Philip Bull,$^c$ David Alonso,$^d$ Stefano Camera,$^e$ Pedro G. Ferreira,$^d$ Gianni Bernardi,$^b$ Roy Maartens,$^{a,f}$ Matteo Viel,$^{g,h}$ Francisco Villaescusa-Navarro,$^{g,h}$ Filipe B. Abdalla,$^{i,j}$ Matt Jarvis,$^{d,a}$ R. Benton Metcalf,$^k$ A. Pourtsidou,$^{k,f}$ Laura Wolz,$^{i}$\\
\llap{$^a$}Department of Physics, University of Western Cape, Cape Town 7535, South Africa\\
\llap{$^b$}SKA SA, The Park, Park Road, Cape Town 7405, South Africa\\
\llap{$^c$}Institute of Theoretical Astrophysics, University of Oslo, 0315 Oslo, Norway\\
\llap{$^d$}Astrophysics, University of Oxford, Oxford OX1 3RH, UK \\
\llap{$^e$}CENTRA, Instituto Superior T\'{e}cnico, Universidade de Lisboa, Lisboa 1049-001, Portugal\\
\llap{$^f$}Institute of Cosmology \& Gravitation, University of Portsmouth,  Portsmouth PO1 3FX, UK\\ 
\llap{$^g$}INAF, Astronomical Observatory of Trieste, 34131 Trieste, Italy\\
\llap{$^h$}INFN, Sezione di Trieste, 34100 Trieste, Italy\\
\llap{$^i$}Department of Physics \& Astronomy, University College London, London, WC1E 6BT, U.K.\\
\llap{$^j$}Department of Physics and Electronics, Rhodes University, PO Box 94, Grahamstown, 6140, South Africa\\
\llap{$^k$}Dipartimento di Fisica e Astronomia, Universit\'{a} di Bologna, 40127 Bologna, Italy\\

E-mail: \email{mgrsantos@uwc.ac.za}}

\abstract{
HI intensity mapping (IM) is a novel technique capable of mapping the large-scale structure of the Universe in three dimensions and delivering exquisite constraints on cosmology, by using HI as a biased tracer of the dark matter density field. This is achieved by measuring the intensity of the redshifted 21cm line over the sky in a range of redshifts without the requirement to resolve individual galaxies. 
In this chapter, we investigate the potential of SKA1 to deliver HI intensity maps over a broad range of frequencies and a substantial fraction of the sky. By pinning down the baryon acoustic oscillation and redshift space distortion features in the matter power spectrum -- thus determining the expansion and growth history of the Universe -- these surveys can provide powerful tests of dark energy models and modifications to General Relativity. They can also be used to probe physics on extremely large scales, where precise measurements of spatial curvature and primordial non-Gaussianity can be used to test inflation; on small scales, by measuring the sum of neutrino masses; and at high redshifts where non-standard evolution models can be probed.
We discuss the impact of foregrounds as well as various instrumental and survey design parameters on the achievable constraints. In particular we analyse the feasibility of using the SKA1 autocorrelations to probe the large-scale signal.
}

\FullConference{
Advancing Astrophysics with the Square Kilometre Array\\
June 8-13, 2014\\
Giardini Naxos, Italy}

\newcommand{\skipthis}[1]{}

\newcommand{\be}{\begin{equation}}
\newcommand{\ee}{\end{equation}}
\newcommand{\bea}{\begin{eqnarray}}
\newcommand{\eea}{\end{eqnarray}}

\begin{document}

\section{Introduction}

The cosmic microwave background (CMB) has been one of the main observational tools for cosmology in recent years. Although basically only giving 2-dimensional information, we were able to constrain the standard cosmological model with great accuracy \citep{2013arXiv1303.5076P}. This "high precision cosmology" is particularly true for the ``vanilla" model with 6 parameters. More parameters or non-standard models can lead to degeneracies and limit the constraining power of the CMB (for instance the $w_0$/$w_a$ non-flat model). The next step towards precision cosmology and exploring novel models will need to use extra information. In particular, due to its huge information content,  measurements of the 3-dimensional large-scale structure of the Universe across cosmic time will be an invaluable tool. One of the most accessible methods to probe this is through large galaxy surveys to trace the underlying dark matter distribution. Several surveys are now under way or in preparation, such as BOSS (SDSS-III), DES, eBOSS, DESI, 4MOST, LSST, and the Euclid satellite. These surveys are based on imaging of a large number of galaxies at optical or near-infrared wavelengths combined with redshift information to provide a 3-dimensional position of the galaxies. 

The new generation of radio telescopes now under construction will provide even larger and deeper radio continuum surveys, capable of detecting galaxies above redshift 3. This is particularly true for the SKA, as discussed in \citet{jarvis}. Although they can provide important constraints on cosmological parameters, they still lack redshift information from the radio that would provide further improvements on the constraints, in particular for the dark energy evolution. A solution is to use the hydrogen 21cm line to provide the redshift information. Telescopes probing the sky between a rest frequency of 1420 MHz and 250 MHz would be able to detect galaxies up to redshift 5. The problem is that this emission line is usually quite weak: at $z=1.5$, most galaxies with a HI mass of $10^9\,M_\odot$ will be observed with a flux density of $\sim1\, \mu$Jy using the HI line. 

In order to obtain ``game changing" cosmological constraints, we showed in \citet{santosSKAHI} that
experiments with sensitivities better than $10\,\mu$Jy over 10 kHz channels will then be required to provide enough galaxies to beat shot noise and become cosmic variance dominated. 
Although ``near term" radio telescopes such as ASKAP and MeerKAT should be able to achieve such sensitivities on deep single pointings, it will require a much more powerful telescope such as SKA Phase 2, to integrate down to the required sensitivity over the visible sky in a reasonable amount of time. This would imply that one would need to wait until then to use radio telescopes for cosmology. 

Galaxy surveys are threshold surveys in that they set a minimum flux above which galaxies can be individually detected. Instead we could consider measuring the integrated 21cm emission of several galaxies in one angular pixel on the sky and for a given frequency resolution. For a reasonably large 3d pixel we expect to have several HI galaxies in each pixel so that their combined emission will provide a larger signal. Moreover we can use statistical techniques, similar to those that have been applied for instance to CMB experiments, to measure quantities in the low signal to noise regime. By not requiring the detection of individual galaxies, the specification requirements imposed on the telescope will be much less demanding. This is what has been commonly called an ``intensity mapping" experiment. It is similar to what is being planned for experiments aimed at probing the Epoch of Reionization (at $z>6$), such as the ones using the radio telescopes LOFAR, MWA and PAPER. By not requiring galaxy detections, the intensity mapping technique transfers the problem to one of foreground cleaning: how to develop cleaning methods to remove everything that is not the HI signal at a given frequency. This in turn also impacts on the calibration requirements of the instrument.

This chapter describes the HI intensity mapping surveys that are feasible with the SKA, listing the assumed telescope specifications as well as the corresponding sensitivity and the different cosmological constraints that can be achieved. A discussion of foreground contamination and calibration requirements is also included. For a summary of the expected cosmological constraints with SKA1 intensity mapping surveys, we refer the reader to section \ref{cosmology} and in particular Fig. \ref{fig:dP}.

\section{The signal}

\begin{figure*}
\begin{center}
\includegraphics[width=0.52\textwidth]{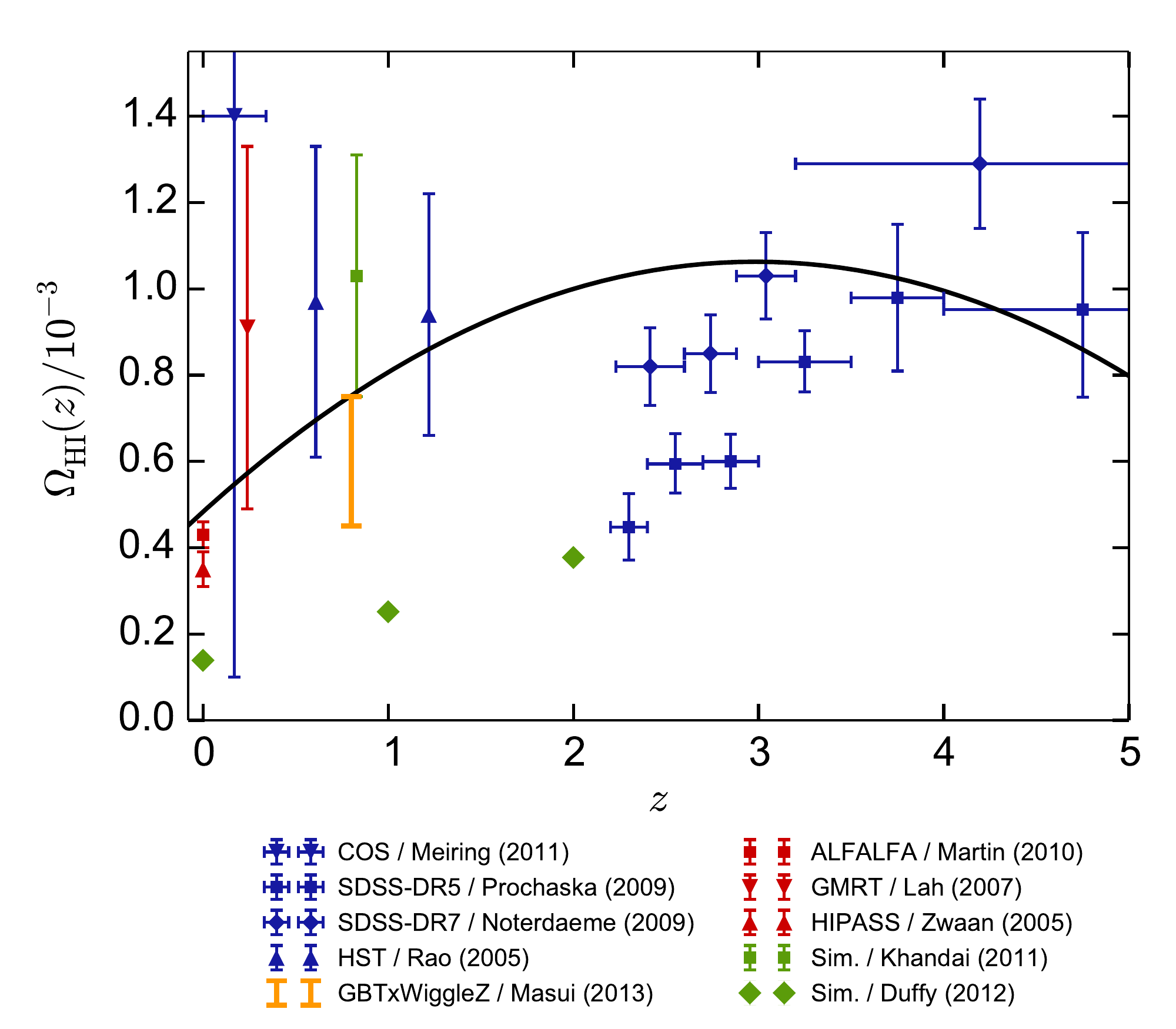}
\includegraphics[trim={0cm -1.5cm 0cm 0cm}, clip,width=0.47\textwidth]{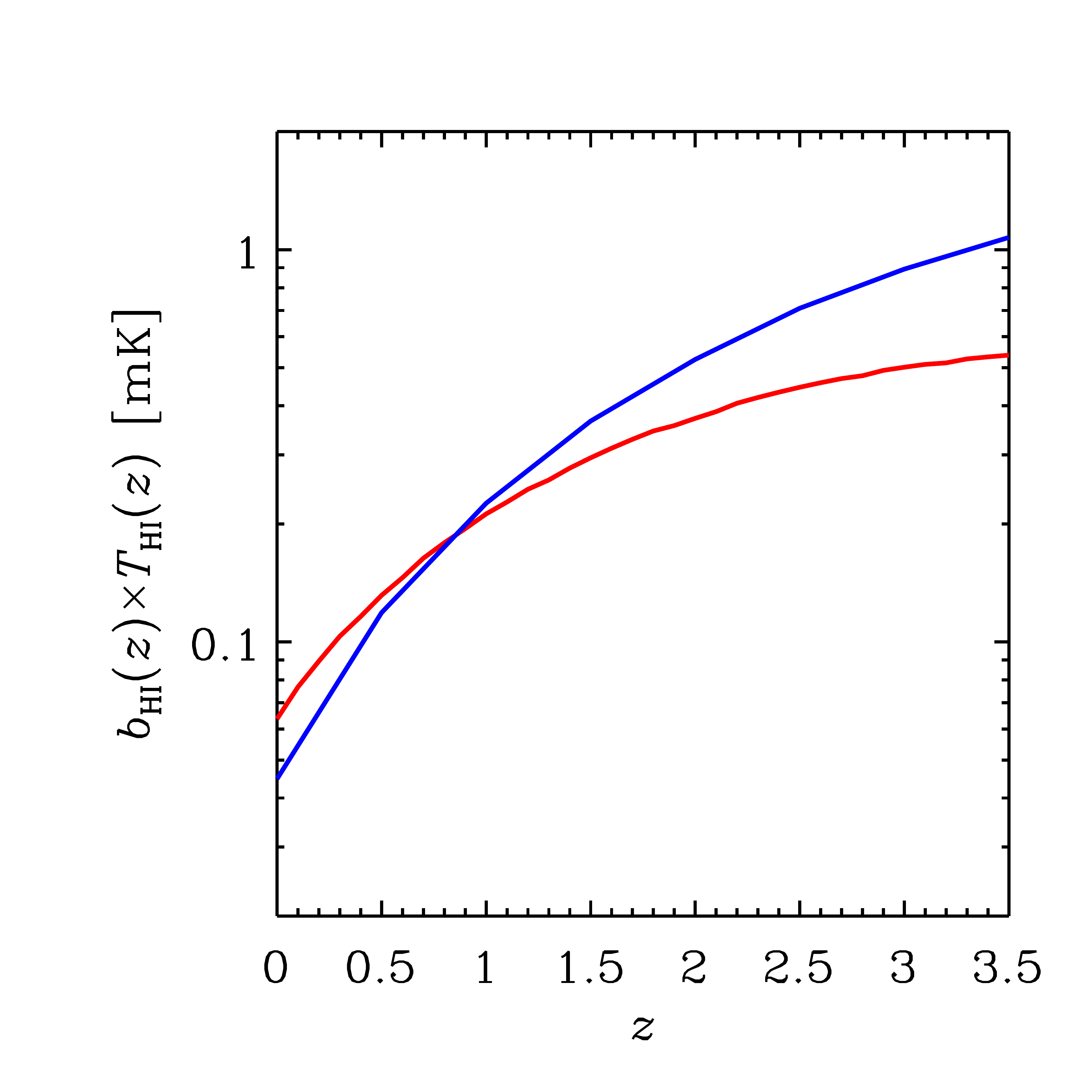}
\caption{Left: Current constraints on the HI density fraction as a function of redshift \citep{Meiring:2011pg, 2009ApJ...696.1543P, 2009A&A...505.1087N, Rao:2005ab, Martin:2010ij, Lah:2007nk, Zwaan:2005cz, 2011MNRAS.415.2580K}, partially based on the compilation in \citet{2012MNRAS.420.2799D}. DLA observations are shown in blue, cross-correlations in orange, other observations in red, and simulations in green. The thick black line shows $\Omega_\mathrm{HI}(z)$ from the fiducial model power law used throughout this chapter. Right: Evolution of the brightness temperature times bias with redshift for the linear (red curve) and our fiducial power-law model (blue).}\vspace{-1.5em}
\label{fig:signal}
\end{center}
\end{figure*}

After reionization, most neutral hydrogen will be found in dense systems inside galaxies, e.g. Damped Lyman-alpha Absorbers (DLAs). In terms of the brightness temperature, the average signal over the sky can be written as:
\be
\overline{T}_{b}(z) \approx 566h\left(\frac{H_0}{H(z)}\right)\left(\frac{\Omega_{\rm HI}(z)}{0.003}\right)(1+z)^2\ {\rm \mu K},
\ee
where the neutral hydrogen density fraction is given by
\be
\Omega_{\rm HI}(z)\equiv (1+z)^{-3} \rho_{\rm HI}(z) / \rho_{c,0},
\ee
$\rho_{\rm HI}(z)$ is the proper HI density and $\rho_{c,0}$ the critical density of the Universe at redshift zero. Figure \ref{fig:signal} shows constraints on $\Omega_{\rm HI}(z)$ from different experiments. At low redshifts, it is measured using 21 cm observations directly from galaxies (except the GBT point which uses intensity mapping). At high redshifts, $\Omega_{\rm HI}$ is estimated by computing the HI associated with Damped Lyman-$\alpha$ systems observed in absorption in quasar spectra. These systems are easy to identify, given their prominent  damping wings in both high-resolution and low-resolution data even at low signal-to-noise, and a HI column density is inferred by Voigt profile fitting. This is in turn easily translated into a value for $\Omega_{\rm HI}$. Present constraints infer a constant $\Omega_{\rm HI}$ at $z=2-4$, while at higher redshift this value is expected to increase, as the Universe is becoming more neutral. For a recent summary of observed trends we refer to \citet{2014arXiv1407.6366P} 

Assuming the signal is linear with respect to the underlying dark matter fluctuations, the total brightness temperature at a given position on the sky and frequency will be
\be
T_{b}(\nu,\Omega) \approx \overline{T}_{b}(z) \Big[1+b_{\rm HI}\delta_m(z)-\frac{1}{H(z)}\frac{dv}{ds}\Big]. 
\ee
The signal will then be completely specified once we find a prescription for the HI density and bias function ($b_{\rm HI}$).
This can be obtained by making use of the halo mass function, $\frac{dn}{dM}$ and relying on a model for the amount of HI mass in a dark matter halo of mass $M$, e.g. $M_{\rm HI}(M)$, so that
\bea
\rho_{\rm HI}(z) &=& \int_{M_{\rm min}}^{M_{\rm max}} dM \frac{dn}{dM}(M,z)\, M_{\rm HI}(M,z),\\
b_{\rm HI}(z) &=& \rho^{-1}_{\rm HI} \int_{M_{\rm min}}^{M_{\rm max}} dM \frac{dn}{dM}(M,z)\, M_{\rm HI}(M,z)\, b(z,M),
\eea
where the halo mass function, $\frac{dn}{dM}$, should be in proper units and $b(z,M)$ is the halo bias. 
%%%%%%%%%%%%%%%%%%%
%%%%%%%%%%%%%%%

Some concerns might arise due to the possible stochastic behaviour of the function $M_{\rm HI}(M)$ or its dependence with the environment (so that it would be a function of position also). However, given the low resolution pixels used in HI intensity mapping experiments, we expect a large number of HI galaxies per pixel, which should average-down any fluctuations and allow us to take the above deterministic relation for the mass function. For example, for the typical scales we are interested in Cosmology, one needs angular/frequency resolutions of around 1 degree and 5 MHz respectively, which translates into a comoving volume of $\sim 10^5$ $\mathrm{Mpc}^3$. In each volume element, we expect a total of around $10^6$ dark matter halos with mass between $10^{8} - 10^{15} M_\odot$, and $\sim\!31,000$ with masses between $5 \times 10^{9}$ and $1 \times 10^{12} M_\odot$ (where the latter range corresponds to halos expected to contain most of the HI mass). This supports our assumption of a position-independent HI mass function due to the averaging over many halos. Some level of stochasticity could still increase the shot noise of the signal, but this is expected to be quite small.

For the mass function, the most straightforward "ansatz" would be to assume that it is proportional to the halo mass -- the constant of proportionality can then be fitted to the available data. Even in this case however, we need to take into account the fact that not all halos contain galaxies with HI mass. Following \citet{Bagla:2009jy}, we can assume that only halos with circular velocities between $30 - 200$ kms$^{-1}$ are able to host HI. This translates into a halo mass through
\be
v_{\rm circ} = 30 \sqrt{1+z} \left(\frac{M}{10^{10}M_{\odot}}\right)^{1/3} ~{\rm kms}^{-1}. 
\ee
Unfortunately this option fails to fit well the HI density measurements at high-z. A more evolved option would be to consider the proportionality to the mass as a function of redshift. This would at least guarantee the fit to the density measurements by construction. Throughout this chapter, we decided to consider instead a simple power law model of the halo mass ($M$):
\be
M_{\rm HI}(M) = A M^\alpha,
\ee
which is independent of redshift.
We found that a value of $\alpha\sim 0.6$ fits both the low $z$ and high $z$ data reasonably well.
This can be seen in figure \ref{fig:signal} (left), that shows the $\Omega_{\rm HI}(z)$ measurements and the evolution obtained from this model (solid line).
The constant $A$ is normalised to the results from \citet{2013MNRAS.434L..46S} at $z\sim 0.8$.  
The right panel shows the redshift evolution for both the linear and power law model of the temperature multiplied by the bias, which is the figure of merit for the strength of the power spectrum used in the forecasts.

\begin{figure*}
\begin{center}
\includegraphics[width=0.46\textwidth]{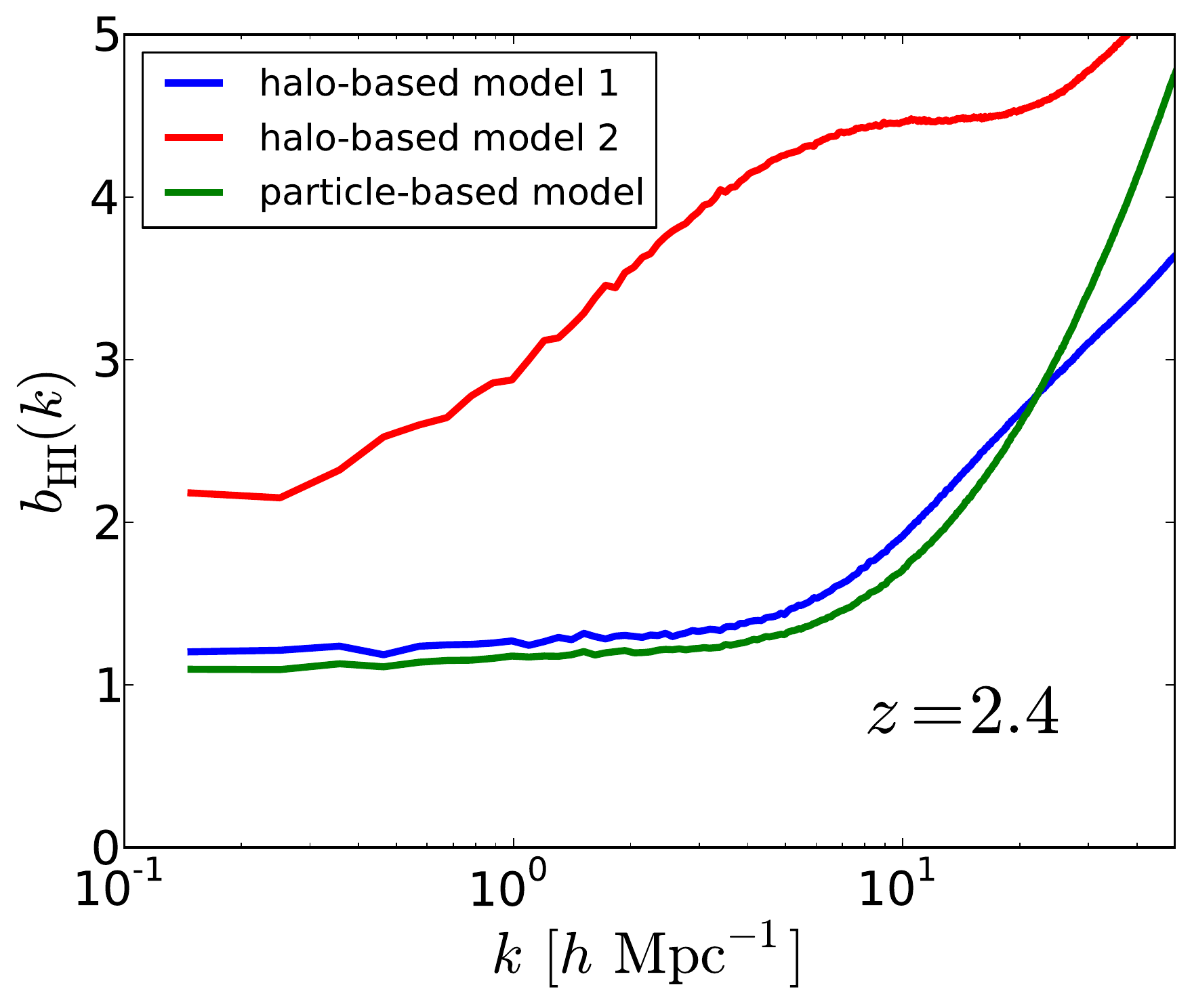}
\includegraphics[trim={-1cm -1.1cm 0cm 0cm}, clip,width=0.48\textwidth]{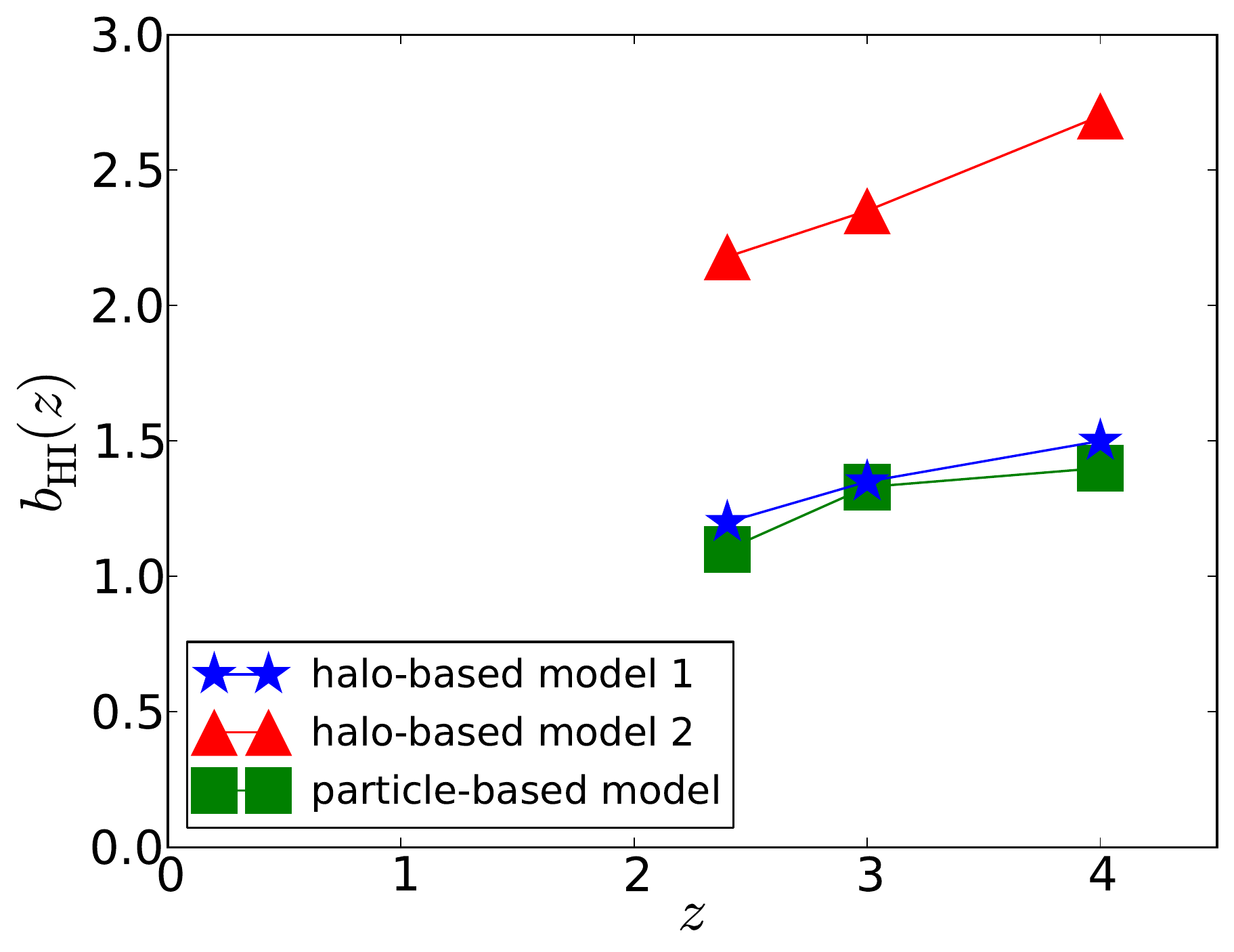}
\caption{HI simulations from \citet{2014arXiv1405.6713V}. Left: HI
  bias as a function of the wave-number at $z=2.4$ for three different
  ways of modelling the HI within and outside haloes.  Halo-based
  model 1 refers to a modification of the modelling performed by
  \citet{Bagla:2009jy}; halo-based model 2 is built in order to fit
  the recent BOSS constraints on the Damped Lyman-$\alpha$ systems;
  particle-based model is inspired by the work of
  \citet{2013MNRAS.434.2645D} which assigns HI both within and outside
  haloes and fits constraints of intergalactic medium data. Note the
  different scale dependence of the three models and the large bias of
  the halo-based model 2. Right: for the same three models we show the
  evolution of the HI bias with redshift in the range probed by the
  simulations.}\vspace{-1.5em}
\label{fig:simsHI}
\end{center}
\end{figure*}

Another issue is whether we can assume that the bias is scale dependent. Again, as long as we restrict ourselves to large scales, this should be a reasonable assumption since we are averaging over many galaxies. Results from simulations show that the bias can be safely assumed constant for $k < 1\ \rm h/Mpc$ at high redshifts (while at $z<1$, it should be safe for $k < 0.1\ \rm h/Mpc$). Note that this bias can also be modelled using a variety of
  relatively simple prescriptions on top of the outputs of large
  volume and high resolution hydrodynamic or N-body simulations
  \citep{2014arXiv1405.6713V}. These models capture the essential
  features of the HI distribution and aim at mimicking more complex
  physical effects (e.g. radiative transfer) by assigning HI to each
  (dark matter or gas) particle of the simulated volume. Although less
  sophisticated than the framework presented in
  \citet{2009ApJ...703.1890O}, these models work remarkably well, allowing us
  to quantitatively address  the scale dependence of the bias. They show that
  at $z>2$ the HI bias steepens for scales $k>1$ Mpc$^{-1}$ in a way which depends on the actual HI modelling as can be seen in Fig. \ref{fig:simsHI}.
  A summary of how uncertainties in the bias signal propagate into the 21 cm HI power spectrum has been recently provided by 
\citet{2014arXiv1407.6366P}.
%%%%%%%%%%%%%%
%%%%%%%%%%%%%%%

%%%%%%%%%%%%%%%%%%%%%%%%%%%%%%%%%%%%%%%%%%
\section{Simulations}
%%%%%%%%%%%%%%%%%%%%%%%%%%%%%%%%%%%%%%%%%%

There exists a number of practical challenges to intensity mapping, for instance the problem of foregrounds (see section \ref{sec:fgs} below), that must be addressed in order to maximize the amount and quality of the scientific information that can be extracted. Quantifying the exact
statistical and systematic uncertainties for a practical experiment analytically can be an
unsurmountable task, and as is now the trend in most cosmological observations, simulations must be
used, which must describe both the cosmological signal we expect to measure and all other
processes (e.g. foregrounds, instrumental effects, etc.) that may have a significant effect
on the recovered data. Ideally we would want these simulations to yield the most realistic
state-of-the-art description possible, however this is not always feasible if a large number of
independent realizations need to be generated in order to quantify the aforementioned
uncertainties. A compromise between computational speed and complexity must be met, so that
enough simulations can be run, while correctly reproducing the relevant physics.

In the case of the cosmological signal for intensity mapping it is possible to accomplish this by
using simplified methods to generate realizations of the matted density field that follow the
correct distribution on large scales. Along these lines, a lot of work has been done in the
last years within the community of galaxy redshift surveys on producing fast but accurate methods
to generate mock simulations of the galaxy distribution \citep{2013JCAP...06..036T,
2014MNRAS.437.2594W}. 
One of the goals of such models is to
  populate dark matter haloes with a realistic distribution of neutral
  hydrogen. To do this, N-body or hydrodynamic simulations
  need to be performed at relatively high resolution to properly
  resolve the smallest haloes that can host HI.  This framework has
  the following features: $i)$ it is simple -- we refer to
  \citet{2014arXiv1405.6713V} for a comprehensive review in which
  three different methods based on \citet{Bagla:2009jy} (halo based)
  and on \citet{2013MNRAS.434.2645D} (particle based); $ii)$ it can be
  easily complemented by a Halo Occupation Distribution model in order
  to cross-correlate HI properties with those of a realistic galaxy
  population; $iii)$ it allows us to address quantitatively the bias
  scale dependence (as we discussed above) and the amount of HI which
  could reside outside haloes; $iv)$ it is bound to reproduce basic
  observable quantities of the HI distribution at high and low
  redshift (like for example Lyman-$\alpha$ forest absorption lines);
  $v)$ it can be translated promptly into realistic observed HI maps,
  once the noise/instrumental properties are known.  
  One recent potentially interesting finding by \citet{2012JCAP...11..059F} with the SDSS-BOSS survey, and using the properties of Damped
  Lyman-$\alpha$ systems, have demonstrated that HI is
  hosted in relatively high mass haloes. If we require the simulations to
  reproduce this, then the HI bias will be larger by a factor of two
  compared to models for which we disregard these observations (this is
  the halo-based model 2 of Fig. \ref{fig:simsHI}).

\begin{figure*}
\begin{center}
  \includegraphics[width=0.48\textwidth]{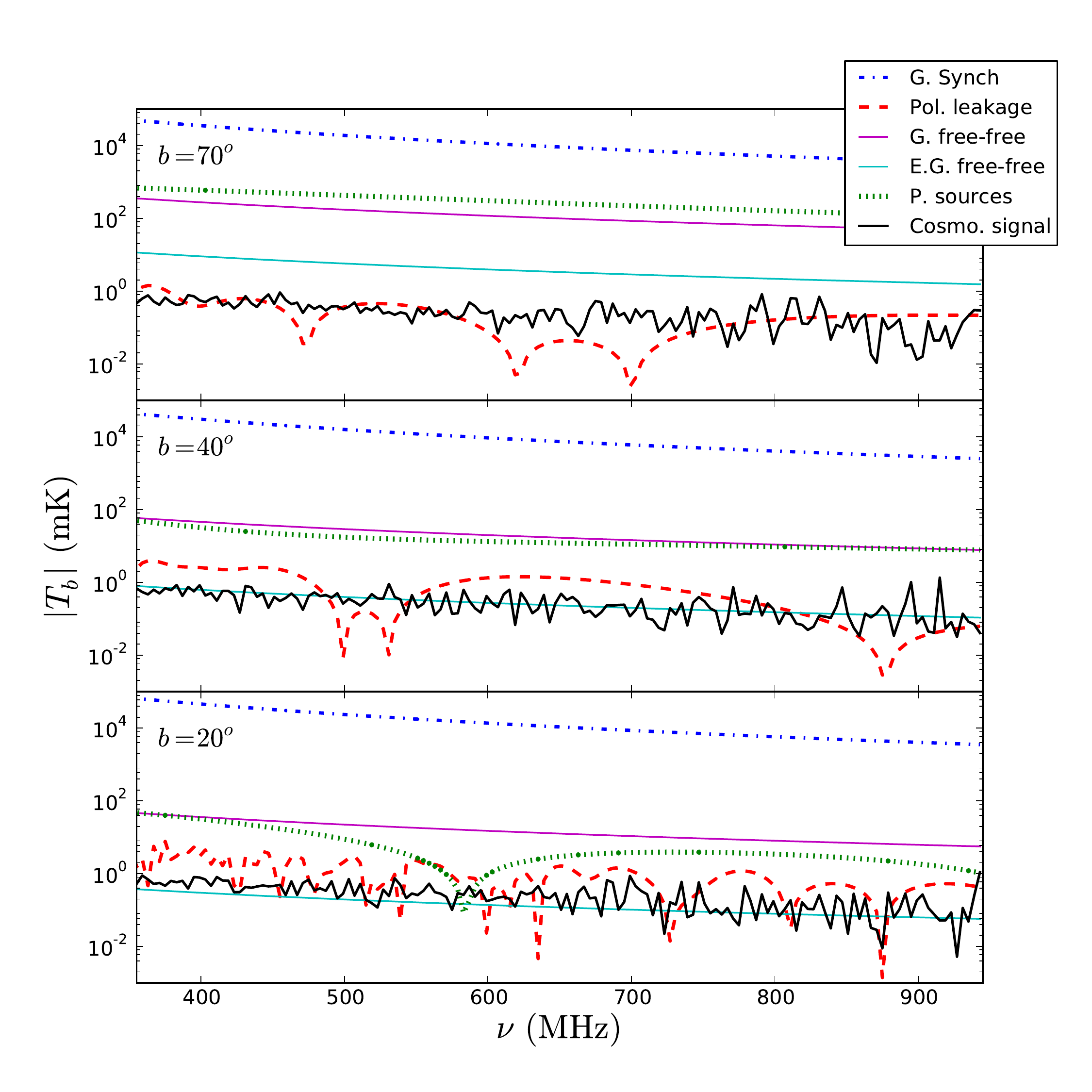}
  \includegraphics[width=0.48\textwidth]{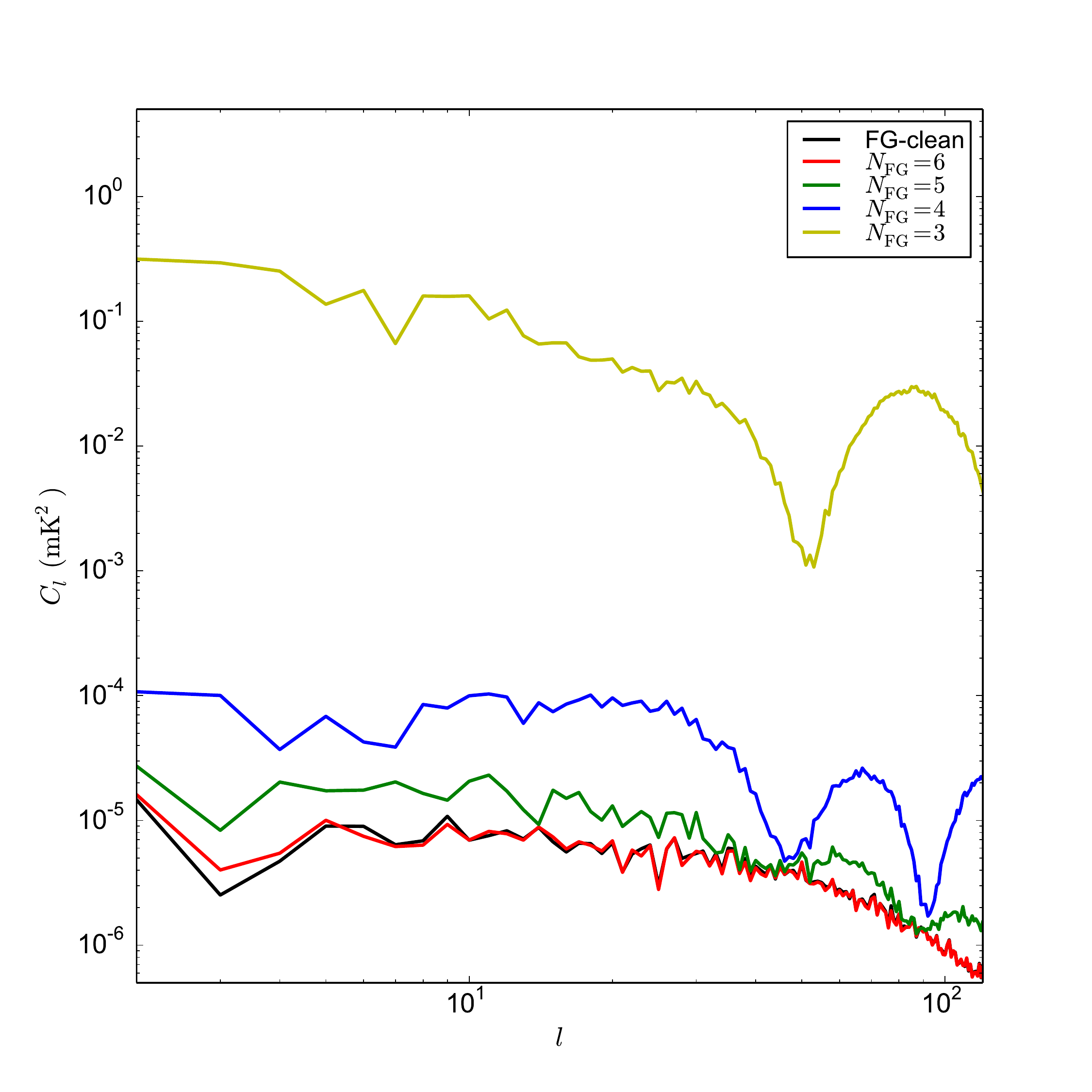}
  \caption{Left: Frequency-dependence of the different foregrounds and the cosmological signal
           at different galactic latitudes (given in the top right corner of each panel),
           according to the simulations of \citet{2014MNRAS.444.3183A}. A 1\% polarization
           leakage was assumed for this simulation. Right: Angular power spectrum of the foreground-cleaned temperature map at 600 MHz for different numbers of foreground degrees of freedom ($N_{\rm FG}$), using a PCA approach. The method converges to the true cosmological signal for $N_{\rm FG}=6$.} 
\label{fig:los_s}\vspace{-1.5em}
\end{center}
%\vspace{-0.5cm}
\end{figure*}

The case of the foregrounds is, however, more complicated. While it is possible to use a few
datasets and certain empirical models to produce conservative realizations of the radio
foregrounds \citep{2008MNRAS.388..247D,2014ApJ...781...57S,2014MNRAS.444.3183A}, the lack of full-sky
multi-frequency data prevents us from developing truly realistic simulations of the radio sky.
This situation will improve in the future as better quality data is obtained by the new radio
observatories, which will occur at the same time as the first intensity mapping observations.
\citet{2014MNRAS.444.3183A} have recently created a publicly available code to generate fast IM mock
observations, including the cosmological signal, foregrounds and some simple instrumental effects.
This code can be used, for example, to study the influence of foreground subtraction on the
recovered cosmological signal, or to analyse the effects of different instrumental configurations
(see figure \ref{fig:los_s}).
%%%%%%%%%%

\section{Foregrounds}\label{sec:fgs}

One of the most important challenges facing HI intensity mapping is the
presence of foregrounds (both galactic and extra-galactic) with amplitudes several orders of
magnitude larger than the signal to be measured. Because the frequency structure as well as other
statistical properties of the foregrounds are significantly different from those of the
cosmological signal, it is not unreasonable to hope that they can be successfully subtracted
\citep{2002ApJ...564..576D,2003MNRAS.346..871O,2005ApJ...625..575S,2006ApJ...648..767M,
2006ApJ...650..529W,2008MNRAS.391..383G,2008MNRAS.389.1319J,2009MNRAS.398..401L,
2009A&A...500..965B,2010A&A...522A..67B,2010MNRAS.409.1647J,2013ApJ...769..154M,
2014MNRAS.441.3271W,2014ApJ...781...57S}. Although a lot of work in terms of simulations and testing cleaning techniques has already been done, we still face huge challenges ahead, in particular if we want to use this signal for high precision cosmology. Increasingly realistic large simulations should be developed to try to test the limitations of the intensity mapping measurements. This should include as many instrumental effects as possible in order to account for possible contamination from the calibration process. Ultimately, we will need to start analysing real data in order to improve and build up our knowledge towards the SKA.

\subsection{Foreground classification}
Foregrounds for intensity mapping can essentially be classified as being extragalactic
(caused by astrophysical sources beyond the Milky Way) or galactic in nature.

The most relevant galactic foregrounds are galactic synchrotron emission (GSE) and
galactic free-free emission. GSE is caused by high-energy cosmic-ray electrons accelerated
by the Galactic magnetic field, and is by far the dominant foreground for HI intensity
mapping, being up to five orders of magnitude larger than the cosmological signal. For the
relevant radio frequencies it should be possible to describe GSE as a power-law in frequency
with a spectral index $\beta\sim-2.8$ \citep{2009ApJ...705.1607D,2013A&A...553A..96D}. Furthermore,
GSE is partially linearly polarised. Its polarised part will be affected by the Faraday effect,
a rotation of the polarization angle caused by the Galactic magnetic field and the optically thick
interstellar medium \citep{1986rpa..book.....R,2009A&A...495..697W}. Since the dependence of this
effect on frequency is quite strong for the frequency range that pertains to intensity mapping, 
any leakage from the polarized part into the unpolarized measurements of the cosmological signal
would generate a very troublesome foreground to subtract \citep{2010MNRAS.409.1647J,
2013ApJ...769..154M}.
On the other hand, free-free emission is caused by free electrons accelerated by ions, and thus
traces the warm ionised medium. As in the case of GSE, free-free emission is predicted to be
spectrally smooth in the relevant range of frequencies \citep{2003MNRAS.341..369D}.
    
Extragalactic radio sources can be classified into two different categories: bright radio
galaxies, such as active galactic nuclei, and ``normal'' star-forming galaxies. The spatial
distributions of these two types should be qualitatively different, the former being less
dominated by gravitational clustering and more by Poisson noise. This has an impact on
the degree of smoothness across frequency of the observed intensity when considering the combined contribution of all the galaxies in a given direction of the sky \citep{2005ApJ...625..575S,
2004ApJ...608..622Z}.

Other possible foreground sources are atmospheric noise, radio frequency
interference and line foregrounds, caused by line emission from astrophysical sources in
other frequencies. Due to the spectral isolation of the 21cm line, together with the expected
low intensity of the most potentially harmful lines (such as OH at $\nu_{\rm OH}\sim1600\,
{\rm MHz}$), the HI signal should be very robust against line confusion.

\subsection{Foreground subtraction}    
The problem of foregrounds has been addressed in the literature mainly within the EoR regime.
The different algorithms that have been proposed to date can be classified into \emph{blind}
\citep{2006ApJ...650..529W,2013MNRAS.434L..46S,2012MNRAS.423.2518C} and \emph{non-blind}
\citep{2011PhRvD..83j3006L,2014ApJ...781...57S,2014arXiv1401.2095S} methods, depending on the
kind of assumptions made about the nature of the foregrounds (e.g. whether only generic
properties such as spectral smoothness and degree of correlation are assumed or whether a
more intimate knowledge of the foreground statistics is required). The poor observational
constraints on the foregrounds in the relevant range of frequencies justifies considering the
use of blind methods. Recently \citet{2014MNRAS.441.3271W} studied the effectiveness of
independent component analysis (in particular the implementation of FastICA, e.g. \citealt{fica1}) for
intensity mapping. By propagating the foreground removal residuals into the cosmological
analysis, they showed that, while foreground
cleaning may induce a residual bias on large angular scales, which could prevent a full analysis
based on the shape of the temperature power spectrum, robust features like the BAO scale should
remain unaffected. This result is reasonable: most relevant foregrounds are (fortunately)
exceptionally smooth and therefore it should be possible to distinguish them from the much
``noisier'' cosmological signal. Any foreground residual will probably be dominated by galactic
synchrotron emmission, which is most relevant on large angular scales.
In the same context, \citet{2014arXiv1409.8667A} studied the efficiency of different
blind cleaning methods (see figure \ref{fig:los_s}), showing that foreground removal can be
successful over a wide range of scales provided the foregrounds are sufficiently smooth, and
that all blind methods yield quantitatively similar results.

In the realm of non-blind methods, the parametric eigenvalue algorithm developed by
\citet{2014ApJ...781...57S} for the CHIME experiment decomposes the data
with help of statistical models for both foregrounds and 21cm signal. This algorithm leaves
minor foreground residuals in the large modes of the power spectrum. Instrumental errors such
as polarization leakage, beam deformations and calibration uncertainties can significantly
affect the foreground removal by mode mixing effects. \citet{2014arXiv1401.2095S} advanced
their description to polarized data
considering a number of instrumental errors in their tests. For future SKA
experiments, detailed studies including varying instrumental settings and the
impact of the residuals on the power spectrum are required in order to minimize
bias on cosmological results. Foreground subtraction for the SKA is treated in detail in
\citet{SKA:IMFG} including realistic simulations.

\subsection{Polarisation leakage}

Although the cosmological signal is unpolarized, sky polarization can represent an additional foreground source due to imperfect calibration. The problem can be described by using the measurement equation formalism \citep{1996A&AS..117..137H,2011A&A...527A.106S} that describes the propagation of the signal through an interferometric array. A pedagogical view can be presented by using the scalar form of the measurement equation \citep{1996A&AS..117..149S}, which relates the measured visibilities to the $(I,Q,U,V)$ Stokes parameters that describe the true sky brightness distribution:
\begin{eqnarray}
V^{pp}_{ij} &=& \frac{1}{2} g^p_i g^{p}_j (I+Q) \nonumber \\
V^{pq}_{ij} &=& \frac{1}{2} g^p_i g^{q}_j (U+iV) \nonumber \\
V^{qp}_{ij} &=& \frac{1}{2} g^q_i g^{p}_j (U-iV) \nonumber \\
V^{qq}_{ij} &=& \frac{1}{2} g^q_i g^{q}_j (I-Q).
\end{eqnarray}
Here, $V$ represents the visibility between antennas $(i,j)$ and polarisations $(p,q)$, e.g. it represents the cross-correlation of the electric fields measured by each antenna (the output of the correlator). The case of "single dish" observations can be simply represented by putting $i=j$. The actual response of the telescope to the input sky is represented here by $g$, usually referred as the "gain" of the system. More instrumental effects can be included in these equations but the simple approach above is enough to show the effect.

If we can calibrate the instrument perfectly, then we can effectively renormalise the gains above (i.e. set $g^p = g^q = 1$) and obtain the measured intensity as $\tilde{I}_{i,j}\equiv V^{pp}_{ij}+V^{qq}_{ij}=I$ (again this is also valid for $i=j$). However, in the presence of calibration errors, there will be an uncertainty in these gains, e.g. $g^{p,q}_{i,j} = 1 + dg^{p,q}_{i,j}$. This will effectively translate into an error in the estimated total intensity:  
$\tilde{I}_{i,j}=I + dI + dQ$ where $dI=\frac{I}{2}(dg^p_i+dg^p_j+dg^q_i+dg^q_j)$ is the usual assumed error due to inaccurate calibration and $dQ=\frac{Q}{2}(dg^p_i+dg^p_j-dg^q_i-dg^q_j)$ is the polarization term that leaks to the total intensity (in the expressions above we assumed small calibration errors). Concentrating in the auto-correlations we have $dI=I (dg^p+dg^q)$ and $dQ=Q (dg^p_i-dg^q_i)$. We then see that even if there is the usual calibration error, the leakage will be zero as long as the error is the same for both polarizations.

Although the polarization leakage is different for different instruments, typical values for leakages are below 1\% and they tend to be reasonably stable over time scales of hours. In this case, the greatest contamination could come from off-axis leakage, i.e. signals entering the telescope from directions other than the pointing one. Their magnitude can be far greater (i.e. up to 30\%), depending upon the observing frequency and their time variability. 
All the algorithms for foreground subtraction rely on the frequency smoothness of the total intensity spectrum. This smoothness is theoretically well motivated \citep{2005ApJ...625..575S, 2011MNRAS.413.2103P, 2014arXiv1404.0887B} but no longer holds for polarization as the Stokes~$Q,U$ parameters are Faraday rotated when the radiation goes through an ionized medium.
The emerging scenario is that intrinsically smooth synchrotron radiation can be contaminated by non-smooth polarized emission due to imperfect calibration. This additional foreground seems to be the limiting factor of current HI mapping measurements \citep{2013MNRAS.434L..46S}. Note however that we can in principle model this contamination from the leakage of faraday rotated polarisation by "looking" at polarised point sources with the telescope.

Unlike the EoR case, both point-like and diffuse Galactic polarized emission may be problematic for intensity mapping at $z \sim 1-2$. The average polarization fraction of extragalactic radio sources is $\sim$5\% at 1.4~GHz \citep{2004MNRAS.349.1267T}) with RM values up to a few tens of rad~m$^{-2}$ at high Galactic latitude where HI intensity mapping is carried out \citep{1981ApJS...45...97S, 2009ApJ...702.1230T}. The properties of Galactic synchrotron polarization are much less known at the frequencies relevant to HI intensity mapping. It is fairly observationally established that the spatial distribution of polarized intensity poorly correlates with total intensity at 1.4~GHz due to small scale structure present in the ionized interstellar medium \citep[i.e.][]{2001ApJ...549..959G,2003MNRAS.344..347B,2011A&A...527A..74S}. Observations of supernova remnants also show that objects further away than a few kpc are completely depolarized at 1.4~GHz, indicating the presence of a polarization horizon beyond which diffuse polarization is no longer observable \citep{2011A&A...527A..74S}. The distance of such polarization horizon decreases at lower frequencies, down to a few hundreds pc at 150-300~MHz \citep{2004A&A...427..169H, 2013ApJ...771..105B}, indicating that relativistic and thermal plasma are co-located in the interstellar medium \citep{1966MNRAS.133...67B}. Typical RM values for Galactic polarization also decrease with decreasing frequencies. Given the complex spatial and frequency properties of Galactic polarization, extrapolations to the frequencies relevant for HI intensity mapping observations are fairly uncertain, although we expect that a significant improvement will happen in the next years due to new surveys.

\section{Experimental considerations}

\subsection{Noise}

Most cosmological applications with HI intensity mapping will rely on the use of statistical quantities, in particular the power spectrum or its equivalent in real space - the 2-point correlation function. To that effect, the noise power spectrum is a prime quantity to access the sensitivity of a given experiment (and associated survey) to detect the cosmological signal. Two main setups can be considered: surveys using single dish observations where the auto-correlation signal from one or more dishes is used, or surveys using interferometers where the cross-correlation signal from the array elements is used. 

\subsubsection{Auto-correlations}
For single dish observations, the noise temperature rms is given by
\begin{equation}
\sigma_T \approx \frac{\lambda^2 T_{\rm sys}}{A_{\rm e} \Delta\Omega \sqrt{2\delta\nu\, t_p}}\approx  \frac{T_{\rm sys}}{\epsilon\sqrt{2\delta\nu\, t_p}}
\end{equation}
where $\Delta\Omega$ is the beam area of the telescope, $A_{\rm e}$ its effective collecting area, $\delta\nu$ the frequency resolution, $t_p$ the observing time per pointing, $\lambda$ the wavelength of the observation and $T_{\rm sys}$ the total system temperature. The factor of $1/\sqrt{2}$ takes into account the fact we have 2 polarisations.
The second approximation takes $\Delta\Omega$ to be the square of the FWHM of the beam and uses a fudge factor $\epsilon\approx 1$ to factor in the efficiency of the telescope. Pointings need to be packed so that the measurement is reasonably continuous across the sky. Usually this means pixels of size $\theta_{\rm B}^2 \approx (\pi/8)(1.3 \lambda/D)^2\ [{\rm sr}]$ or smaller. Given a total observing time $t_{\rm tot}$, the time per pointing is then $t_p=t_{\rm tot} (\theta_{\rm B})^2/S_{\rm area}$ where $S_{\rm area}$ is the survey area. The 3D noise power spectrum is just $P_N=\sigma_T^2\,V_{\rm pix}$, where $V_{\rm pix} = (r \theta_{\rm B})^2 \times (y\delta\nu)$ is the 3D comoving volume of each volume element, $r$ is the comoving distance to the redshift of the signal and $y=c H(z)^{-1}(1+z)^2/\nu_{21}$. Therefore, the $\theta_{\rm B}^2$ cancels in the power spectrum and we finally get
\be
P_N=r^2 y \frac{T_{\rm sys}^2 S_{\rm area}}{2 \epsilon^2 t_{\rm tot}}.
\ee
As we can see, the dish size drops out of the final expression - it is only relevant for the angular resolution which should match what is required for the signal we are trying to measure (to summarise, the dish collecting area and beam size are connected and will cancel out and the way we "pack" the beams for mosaicking is connected to the assumed pixel resolution in the map and will cancel out in the power spectrum). For a fixed total observation time, the survey area should also be chosen to match the required angular scales. If the noise power spectrum is similar to the signal, one also gains by increasing the survey area since this will increase the number of independent measurements for a given scale.

If we have $N_d$ dishes, the combined noise power spectrum will be $P_N = r^2 y T_{\rm sys}^2 S_{\rm area} / (2 \epsilon^2 N_d t_{\rm tot})$. For a single dish with $N_b$ beams, we can cover the same sky area in less time, so that the noise power spectrum will go as $P_N=r^2 y T_{\rm sys}^2 S_{\rm area}/(2 \epsilon^2 N_d N_b t_{\rm tot})$. With Phased Array Feeds (PAFs) the situation is slightly more complicated as the feeds are packed in order to allow for a large number of beams. This will imply some amount of beam overlap below a given critical frequency 
%%%%%%
\begin{equation}
P_N = r^2 y \frac{T_{\rm sys}^2 S_{\rm area}}{2 \epsilon^2 N_d N_b t_{\rm tot}}
\times\left\{ 
  \begin{array}{l l}
    1 & \quad \nu > \nu_\mathrm{crit} \\[1em]
    \left(\frac{\nu_\mathrm{crit}}{\nu}\right)^2 & \quad \nu \le \nu_\mathrm{crit}
  \end{array} \right. ,
\end{equation}
where $\nu_\mathrm{crit}$ is the PAF critical frequency. Note that $S_{\rm area} > N_b \theta_B^2$ (with $\theta_B^2 \propto 1/\nu^2$ as usual).

\subsubsection{Interferometer}

For observations with interferometers, we start by considering the noise rms in the uv plane for a "uv" pixel of size $(\Delta u)^2$:
\begin{equation}
\sigma_T(\mathbf{u},\nu) = \frac{\lambda^2 T_{\rm sys}}{A_e\sqrt{2 \delta\nu\, n(\mathbf{u})\, (\Delta u)^2 t_p}},
\end{equation}
where $A_e$ is the effective collecting area of one element (dishes or stations), $t_p$ the time per pointing and $n(\mathbf{u})$ is the average number density of baselines (averaged over a 24h period), usually only a function of $|\mathbf{u}|$.

For interferometers, we are going to assume that mosaicking different pointings will not allow the recovery of angular scales larger than the telescope field of view (the primary beam), which is basically set by the size of the array elements, e.g. $\theta_{\rm B}^2\sim \lambda^2/D^2\ [sr]$, where $D$ is the dish or station diameter. Usually, the total observing time $t_{\rm tot}$ and time per pointing $t_p$ are then the same. On the other hand, we can use different pointings to increase the number of independent measurements on scales smaller than the telescope field of view. The time per pointing should then be decreased as $t_p=t_{\rm tot}/N_p$, where $N_p$ is the number of pointings. 
The 3d noise power spectrum is then given by
\be
P_N(\mathbf{k},\nu)= \frac{\lambda^4 r^2\,y T_{\rm sys}^2 N_p}{2 A_e^2 n(u) t_{\rm tot}} = \frac{\lambda^4 r^2\,y T_{\rm sys}^2 S_{\rm area}}{2 A_e^2 \theta_{\rm B}^2(\nu) n(u) t_{\rm tot}}.
\ee
with $u=r(z) k_\perp/(2\pi)$ and $N_p\equiv S_{\rm area}/\theta_{\rm B}^2$.
Note that if we assume that $n(u)$ is constant on the uv plane between some minimum $D_{\rm min}$ and maximum $D_{\rm max}$ baseline, then we can write
\begin{equation}
n(u)=\frac{N_a(N_a-1)\lambda^2}{2\pi(D_{\rm max}^2-D_{\rm min}^2)}.
\end{equation}
where $N_a$ is the number of elements of the interferometer contributing to that baseline range. The expression follows by noting that the integration over $n(u)$ should give the total number of baselines. In the analysis below, we considered the full $n(u)$ distribution\footnote{files with n(u) for the different telescope setups are available at https://gitorious.org/radio-fisher/bao21cm.git}.

For PAFs, each of the beams will be cross-correlated with the corresponding beam from a different dish (e.g., the field of view of the interferometer is set by the size of one of these beams). The total number of beams per dish, $N_b$, will allow to survey a target area more quickly and thus increase $t_p=t_{tot} N_b \theta_{\rm B}^2 / S_{\rm area}$. However, we need to consider again the beam overlap below the critical frequency so that
%%%%%%
\begin{equation}
P_N = \frac{\lambda^4 r^2\,y T_{\rm sys}^2 S_{\rm area}}{2 A_e^2 N_b n(u) \left[\theta_{\rm B}(\nu_\mathrm{crit})\right]^{2} t_{\rm tot}}
\times \left\{ 
  \begin{array}{l l}
    \left(\frac{\nu}{\nu_\mathrm{crit}}\right)^2 & \quad \nu > \nu_\mathrm{crit} \\[1em]
    1 & \quad \nu \le \nu_\mathrm{crit}
  \end{array} \right. .
\end{equation}
Note that the collecting area of one element of a dish array, $A_e$ can be written as $A_e \approx  \pi D^2\epsilon/4$ with $\epsilon \lesssim 1$. 

If we consider aperture arrays such as what is used for SKA-LOW, the situation is slightly different. In that case, above a given critical frequency, the area of a station will go as $1/\nu^2$, being constant below that (when the array becomes dense). On the other hand, the array beam $\theta_{\rm B}$ should go as $1/\nu$ at any frequency (set by the size of the array) so that
%%%%%%
\begin{equation}
P_N = \frac{\lambda^4 r^2\,y T_{\rm sys}^2 S_{\rm area}}{2 \left[A_e(\nu_\mathrm{crit})\right]^2 \theta_{\rm B}^2(\nu) N_b n(u) t_{\rm tot}}
\times \left\{ 
  \begin{array}{l l}
    \left(\frac{\nu}{\nu_\mathrm{crit}}\right)^4 & \quad \nu > \nu_\mathrm{crit} \\[1em]
    1 & \quad \nu \le \nu_\mathrm{crit}
  \end{array} \right. ,
\end{equation}
where we are already considering the possibility of multiple beams ($N_b$) with SKA-LOW.

\subsubsection{Total error}

Computation of the total error in the measurement of the power spectrum at a given scale, $P(k)$, will require knowledge of both the noise and the number of independent modes used to measure that scale, since the error is $\sim (P(k)+P_N(k))/\sqrt{N_{\rm modes}}$. These number of modes will be related to the volume of the survey and optimisation will depend on the balance between reducing noise and increasing the number of modes. Moreover, issues such as the large $k$ cutoff along the angular and frequency directions need to be factored in due to the resolution of the experiment. All these details have been taken into account in the forecasting and fully described in \citet{2014arXiv1405.1452B}.

\subsection{Target scales}

The scales we want to probe for cosmology will impose requirements on the telescope specifications (or the other way around). In terms of probing baryon acoustic oscillations (BAO - see \citealt{bull}), the relevant scales can be translated to:
\begin{itemize}
\item
Angular scales between 30 arcminutes and 4 degrees.
\item
Frequency scales between 2 MHz and 35 MHz
\item
Surveys covering large areas of the sky are required in order to increase the statistical detection.
\item
A large bandwidth is required in order to maximize the redshift range covered. Ideally: $\gtrsim 350$ MHz $(0 < z \lesssim 3)$. Though a split in bands is probably required: the low frequency part will be important for large scale physics and "non-standard" cosmological constraints while the high frequency/low $z$ part will be useful to probe the more standard "vanilla" models. Since there will be other surveys probing this low $z$ region we suggest to focus on the $< 1000$ MHz region if a split in bands needs to be made.
\end{itemize}
For GR corrections and non-Gaussianity, we are interested on ultra-large scales, e.g. modes $k\lesssim 0.01$ Mpc$^{-1}$. This should allow high precision measurements to be made of the turn over scale in the power spectrum. These scales will correspond to angular sizes $> 10$ deg (at $z\sim 1$) and frequency intervals of order 100 MHz which should be easily achievable (but note that the foreground cleaning scale will require even larger bandwidths). These angular requirements mean that we will not be able to use interferometers to probe these large scales since their primary beam is usually smaller than this (and we are assuming that mosaicking cannot be used to recover the large scales with the interferometer).

The signal we are looking for fluctuates both in frequency and across the sky. We are not looking for the average intensity. In table \ref{table:rms} we show the expected signal rms fluctuations for scales related to the dish FoV and a couple of frequency intervals.
%%%%%%%%%%%%%%%
\begin{table}
\begin{center}
\begin{tabular}{@{}|cccccc|}
\hline 
Redshift  &	Frequency  &  Average Signal  &	Angular scale & Signal rms               &        Signal rms\\
               &       (MHz)     &     ($\mu \rm K$)     &     (deg)           & ($\mu \rm K$ - 1 MHz)  &  ($\mu \rm K$ - 10 MHz)\\
\hline 
0.5 &	947 &	 156.5  & 1.4 & 36.2 & 29.0\\
1.0 &	710 &  254.8  & 1.9 & 29.8 & 24.7\\
1.5 &	568 &  351.0  & 2.4 & 25.9 & 21.5\\
2.0 &	473 &  427.8  & 2.8 & 23.2 & 19.1\\
 \hline
\end{tabular} 
\end{center}
\caption{Average of the signal as a function of redshift/frequency as well as the signal rms (square root of the variance) for scales relevant for BAO ($\sim 150$ Mpc) assuming single dish observations with $\sim 15$ m dishes. The variance of the signal is calculated for a 3d smoothing window corresponding to the given angular and frequency scale (frequency in parentheses).}
\label{table:rms}
\end{table}

\subsection{Current and planned experiments}

First attempts at using intensity mapping have been promising, but have highlighted the challenge of calibration and foreground subtraction. The Effelsberg-Bonn survey \citep{2011AN....332..637K} has produced a data cube covering redshifts out to $z=0.07$, while the Green Bank Telescope (GBT) has produced the first (tentative) detection of the cosmological signal through IM by cross-correlating with the WiggleZ redshift survey \citep{Chang:2010jp,2013MNRAS.434L..46S,2013ApJ...763L..20M}. As probes to constrain cosmological parameters these measurements are, as yet, ineffective, but they do point the way to a promising future.

As described above, we can divide the intensity mapping experiments into two types: single dish surveys and interferometers. In single dish surveys (e.g. using auto-correlations) each pointing of the telescope gives us one single pixel on the sky (though more dishes or feeds can be used to increase the field of view). This has the advantage of giving us the large scale modes by scanning the sky. Since brightness temperature is independent of dish size we can achieve the same sensitivity with a smaller dish although that will in turn limit the angular resolution of the experiment (a 30 arc min resolution at $z\sim 1$ would require a dish of about 50 m in diameter). One example is the GBT telescope as described above. BINGO \citep{2013MNRAS.434.1239B} is a proposed 40m multi-receiver single-dish telescope to be situated in South America and aimed at detecting the HI signal at $z\sim 0.3$.

Interferometers basically measure the Fourier transform modes of the sky. They have the advantage of easily providing high angular resolution as well being less sensitive to systematics that can plague the auto-correlation power. On the other hand, the minimum angular scale they can probe is set by their shortest baseline which can be a problem when probing the BAO scales. One example of a purpose built interferometer for intensity mapping is CHIME, a proposed array, aimed at detecteding BAO at $z\sim 1$, made up of $20 \times 100$m cylinders, based in British Columbia, Canada. The pathfinder has 2 half-length cylinders, and the full experiment has 5 \citep{chime}.

The next generation of large dish arrays can also potentially be exploited for HI intensity mapping measurements. Such is the case of MeerKAT and ASKAP. However, these interferometers do not provide enough baselines on the scales of interest (5m to 80m) so that their sensitivity to BAO will be small. The option is to use instead the auto-correlation information from each dish, e.g. make a survey using the array in single dish mode. The large number of dishes available with these telescopes will guarantee a large survey speed for probing the HI signal.
The great example of this approach will be SKA1, the first phase of the SKA telescope, to be built in 2018. 
A HI intensity mapping survey will turn SKA phase 1 into a state of the art cosmological probe and we discuss its use in the next sections.
%%%%%%%%%%%%%%%
%\begin{figure}[]
% \vspace*{-2.0 cm}
%\begin{center}
% \includegraphics[width=2.8in]{pub-resolution-bao.pdf} 
% \vspace*{-1.0 cm}
% \caption{Redshift evolution of the minimum/maximum transverse scales (filled regions) for illustrative interferometer (blue) and single-dish (red) experiments. The BAO are plotted for comparison. The dishes have diameter $D_\mathrm{dish} = 15$m, the min./max. interferometer baselines are $D_\mathrm{min} = 15$m and $D_\mathrm{max} = 1000$m, and the survey has bandwidth $\Delta \nu = 600$ MHz and area $S_\mathrm{area} = 25,000$ sq. deg. The shaded grey region denotes superhorizon scales, $k < k_H = 2 \pi / r_H$.}
 %  \label{f4}
%\end{center}
%\end{figure}
%%%%%%%%%

\section{Surveys with the SKA}

In terms of HI intensity mapping surveys with SKA, several factors need to be considered:
\begin{itemize}
\item
SKA phase 1 (SKA1) is planned to have 3 different instruments and all of them can in principle provide an intensity mapping survey at redshifts below 5 (after reionization). This includes SKA1-LOW which is planned to operate at frequencies below 350 MHz (so $z\gtrsim 3$). SKA1-MID and SKA1-SUR are on the other hand planned to work down to 350 MHz, although the deployment of the required bands might not happen at the same time.
\item
Both SKA1-MID and SUR can be used in "single dish mode" where the auto-correlations are used to probe the large cosmological scales, or in "interferometer mode" better at resolving the smaller scales. For SKA-LOW we will only consider the interferometer mode. In principle one can also consider a survey with SKA-LOW using the auto-correlation of the beam from each station (from the beam-former). This could be useful for probing large scales with a full sky survey but wouldn't be optimised for BAO since at $\sim 350$ MHz the beam would be around 3 deg$^2$. Other option would be to consider the dipoles as the correlation elements instead of stations. These are possibilities that need to be further explored, but since SKA-LOW is more focused on high redshifts where reionization effects need to be factored in, for the current analysis we decided to concentrate only on the standard interferometer case as an example for SKA-LOW.
\item
Both SKA1-MID and SUR will have several bands and, according to the current design, we will need to use two bands to cover the full redshift range from $z=0$ to $z=3$.
\item
The SKA is probably going to be built in 3 phases (or even 4 if we consider the SKA precursors, MeerKAT and ASKAP). We should therefore consider in the analysis two more phases besides SKA1: SKA Phase 0 - An "early science" phase of deployment for each SKA1 component (SKA1-LOW, SKA1-SUR, SKA1-MID), where sensitivity has grown to about $50\%$ of its fully specified level and the full SKA (SKA Phase 2), with 10x the sensitivity and 20x the field of view of SKA1.
\end{itemize}

The baseline design for SKA1 is described in \citet{dewdney2013ska1} and further updated in \citet{braun2014}.
To summarise, SKA1-MID will comprise of 254 single pixel feed dishes (including MeerKAT 64 dishes), to be built in South Africa, and 96 dishes (including ASKAP 36 dishes)  fitted with 36 beam PAFs to increase the field of view to be set in Australia. SKA1-LOW consists of 911 aperture array stations each with 35 m diameter. Following the details in table 1 of \citet{santosSKAHI} we list in table \ref{tab:surveys} the different surveys that are considered in this chapter.
%%%%%
\begin{table*}[t]
%\begin{center}
{\renewcommand{\arraystretch}{1.0} 
\tabcolsep=0.09cm
{\small
\begin{tabular}{|c|c|c|c|c|c|c|c|c|c|}
\hline
Telescope $^{(a)}$ & Band [MHz] $^{(b)}$& z & Target freq.  & $T_\mathrm{inst} \, [\mathrm{K}]$ & $N_d$ &
 $D \, [\mathrm{m}]$ $^{(c)}$ & A$_{\rm eff}\, [\mathrm{m^2}]$$^{(d)}$  &
  $\theta_B^2$ $[\mathrm{deg}^2]$ $^{(e)}$  &  $N_b$\\
\hline
SKA0-MID & 900 - 1760 & (0) - 0.58 & 1310 [MHz] & 20 & 127 & 15 &  140 &  0.51 & 1   \\
SKA0-SUR $^{(f)}$ & 650 - 1800 $^{(g)}$ & (0) - 1.19 & 1300 & 30 & 48 &   15 & 140  &  0.51 & 36 $^{(h)}$ \\
\hline
SKA1-MID & 350 - 1050 & 0.35 - 3.06 & 700 & 28 & 254 & 15 &  140  &  1.78 & 1  \\
SKA1-MID & 900 - 1760 & (0) - 0.58 & 1310 & 20   & 254 & 15 & 140  &  0.51 & 1  \\
SKA1-SUR & 350 - 900 $^{(g)}$ & 0.58 - 3.06 & 710 & 50 & 60  & 15 &  140 &  1.71 & 36  $^{(h)}$\\
SKA1-SUR $^{(f)}$ & 650 - 1800 $^{(g)}$ & (0) - 1.19 & 1300 & 30 & 96  & 15 & 140  &  0.51 & 36 $^{(h)}$ \\
\hline
SKA1-LOW & 50 - 350  $^{(i)}$  & 3.06 - 27  & 110 & 40 & 911  & 35 & 925 $^{(j)}$  & 28 & 3 $^{(k)}$\\
\hline
SKA2 $^{(l)}$& 300 - 1000  & 0.42 - 3.73 & 500 & 15 & 4000 & 10 & 63  & 8.0 & 10 \\ 
\hline

\end{tabular}}
}
\caption[x]{Telescope/survey configurations. For frequency dependent quantities, the values are calculated at the indicated target frequency. In order to compare between different setups, a total of 10,000 hours is assumed for each survey. Both single dish and interferometer data are considered for each survey when possible. Survey area is chosen to optimise detection at the required cosmological scale. For all surveys we take the total temperature to be $T_{\rm sys}=T_{\rm rcvr}+T_{\rm sky}$ with $T_{\rm rcvr}=0.1 T_{\rm sky} + T_\mathrm{inst}$ and $T_{\rm sky}\approx 60 \left(300\, {\rm MHz}/\nu\right)^{2.55}$ K.\\
{\bf Notes:} {\bf (a)} MID and SUR telescopes are already assumed to include MeerKAT (64 dishes) and ASKAP (36 dishes) respectively.
{\bf (b)} For MID and SUR the largest band of the combined telescopes is indicated assuming that outside the overlapping band only the corresponding dishes are used in the sensitivity calculations.
{\bf (c)} Diameter of dish or station.
{\bf (d)} Effective collecting area of the dish or stations.
{\bf (e)} Primary beam or instantaneous field of view of the telescope at the target frequency - assumed to go as $1/\nu^2$ unless stated otherwise. For the combined telescopes, the smallest beam of the two telescopes is used.
{\bf (f)} Assuming that all ASKAP PAFs will be replaced to meet the SKA1-SUR band and
instrument temperature of 30K. 
{\bf (g)} Only 500 MHz instantaneous bandwidth.
{\bf (h)} The combined PAF beam ($N_b\times \theta_B^2$) is assumed constant below the target (critical) frequency as explained in the text.
{\bf (i)} Note that band$\times\ N_b$ is fixed at 300 MHz.
{\bf (j)} Assumed to be constant below the target frequency and going as $1/\nu^2$ above it.
{\bf (k)} Beams can point in different directions so no overlap is assumed.
{\bf (l)} Values here are completely indicative. We choose to consider an interferometer type experiment, targeting the high ($z>1$) redshift region and a FoV large enough for BAO scales. We only fix the sensitivity to be about 10 times what is assumed for SKA1 and the total field of view (including number of beams) to be about 20 times SKA1 at the target frequency.}
\label{tab:surveys}
\end{table*}
%%%%%

%%%%%%%%%%%%%%%%%%%%%%%%%%%%%%%%
\section{Cosmological constraints}
%%%%%%%%%%%%%%%%%%%%%%%%%%%%%%%%
\label{cosmology}

\subsection{Dark energy and spatial curvature}

Surveys of large-scale structure are a rich source of information about the geometry and expansion history of the Universe. The baryon acoustic oscillations (BAO) are a preferred clustering scale imprinted in the galaxy distribution, originating from the time when photons and baryonic matter were coupled together in the early Universe. By using them as a statistical `standard ruler', one can obtain constraints on the expansion rate, $H(z)$, and (angular) distance-redshift relation, $D_A(z)$, as functions of redshift, as has been done successfully with recent large galaxy redshift surveys such as BOSS and WiggleZ. Measuring these functions is vital for testing theories of dark energy which seek to explain the apparent acceleration of the cosmic expansion, as they constrain its equation of state, $w = P/\rho$, and thus its physical properties. Shedding light on the behaviour of dark energy -- especially whether $w$ deviates from $-1$ and whether it varies in time -- is one of the foremost problems in cosmology.

\begin{figure}[t]
\centering{
\includegraphics[width=0.65\columnwidth]{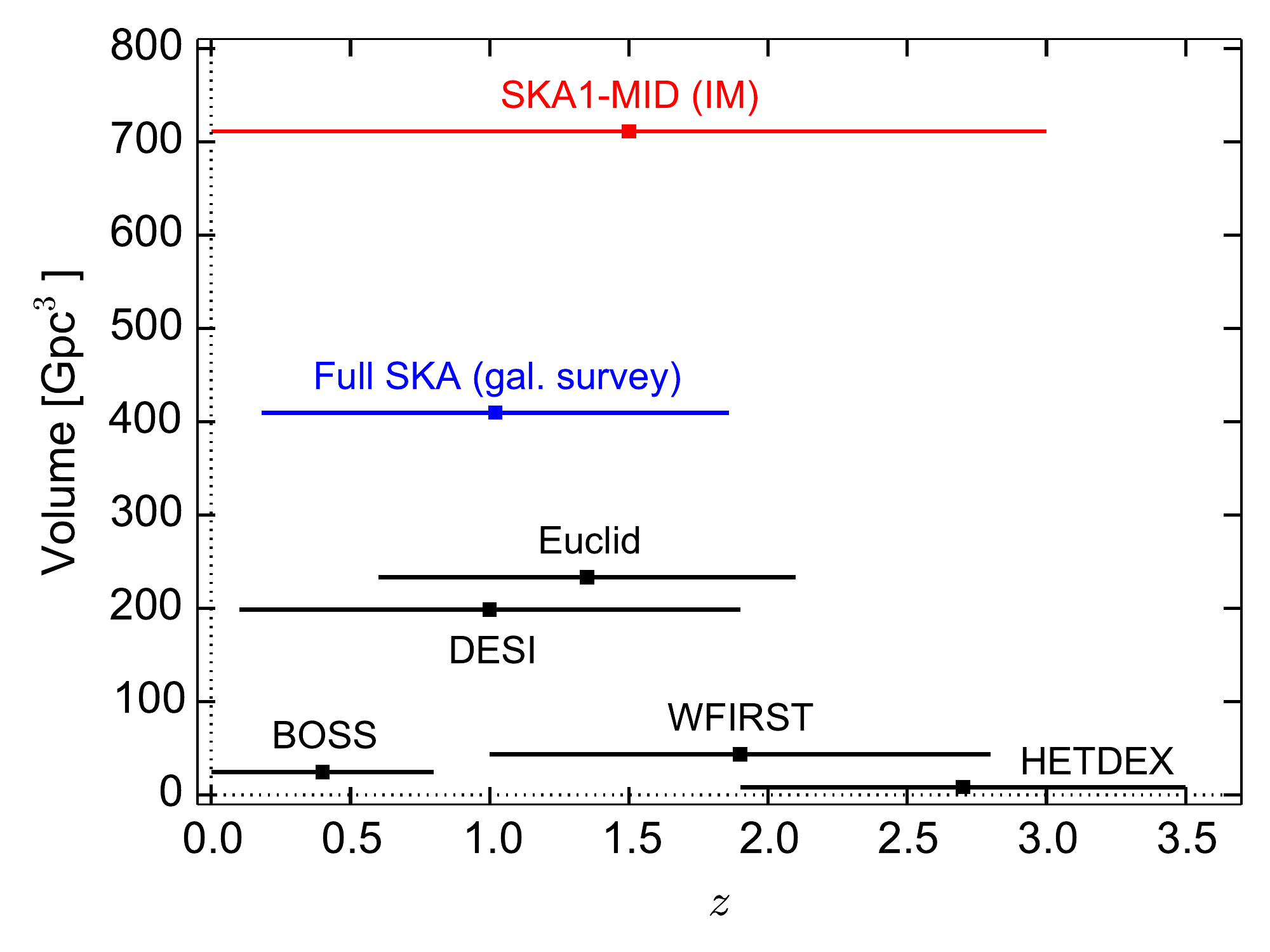}
\vspace{-1.5em}\caption{Survey volumes and redshift range for various current and future surveys (volume calculated at the central redshift).} \vspace{-1.5em}
\label{fig:vol} }
\end{figure}

To precisely measure the BAO feature in the matter correlation function, which appears as a `bump' at comoving separations of $r \approx 100 h^{-1}$ Mpc, one needs to detect many galaxies (in order to reduce shot noise), and to cover as large a survey volume as possible (in order to reduce sample variance). Intensity mapping has a few major advantages over conventional galaxy surveys for this task. IM surveys can map a substantial fraction of the sky with low angular resolution in a short period of time. Combined with the wide bandwidths of modern radio receivers, this makes it possible to cover extremely large survey volumes and redshift ranges in a relatively short time, helping to beat down sample variance (see Fig. \ref{fig:vol}). 
%%%%%%%
%%%%%%%%%%

\begin{figure*}
\begin{center}
  \includegraphics[width=0.495\textwidth]{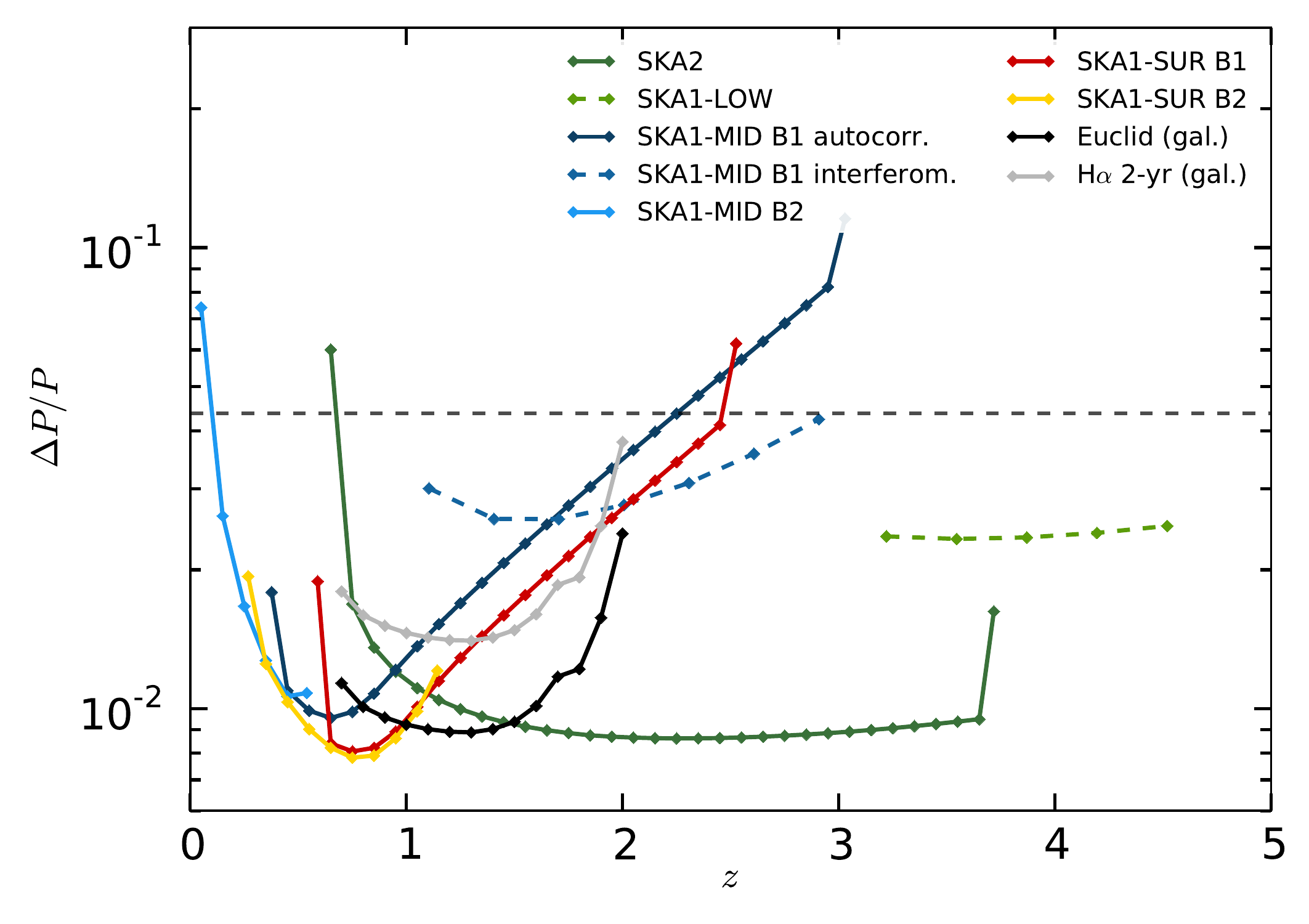}
  \includegraphics[width=0.495\textwidth]{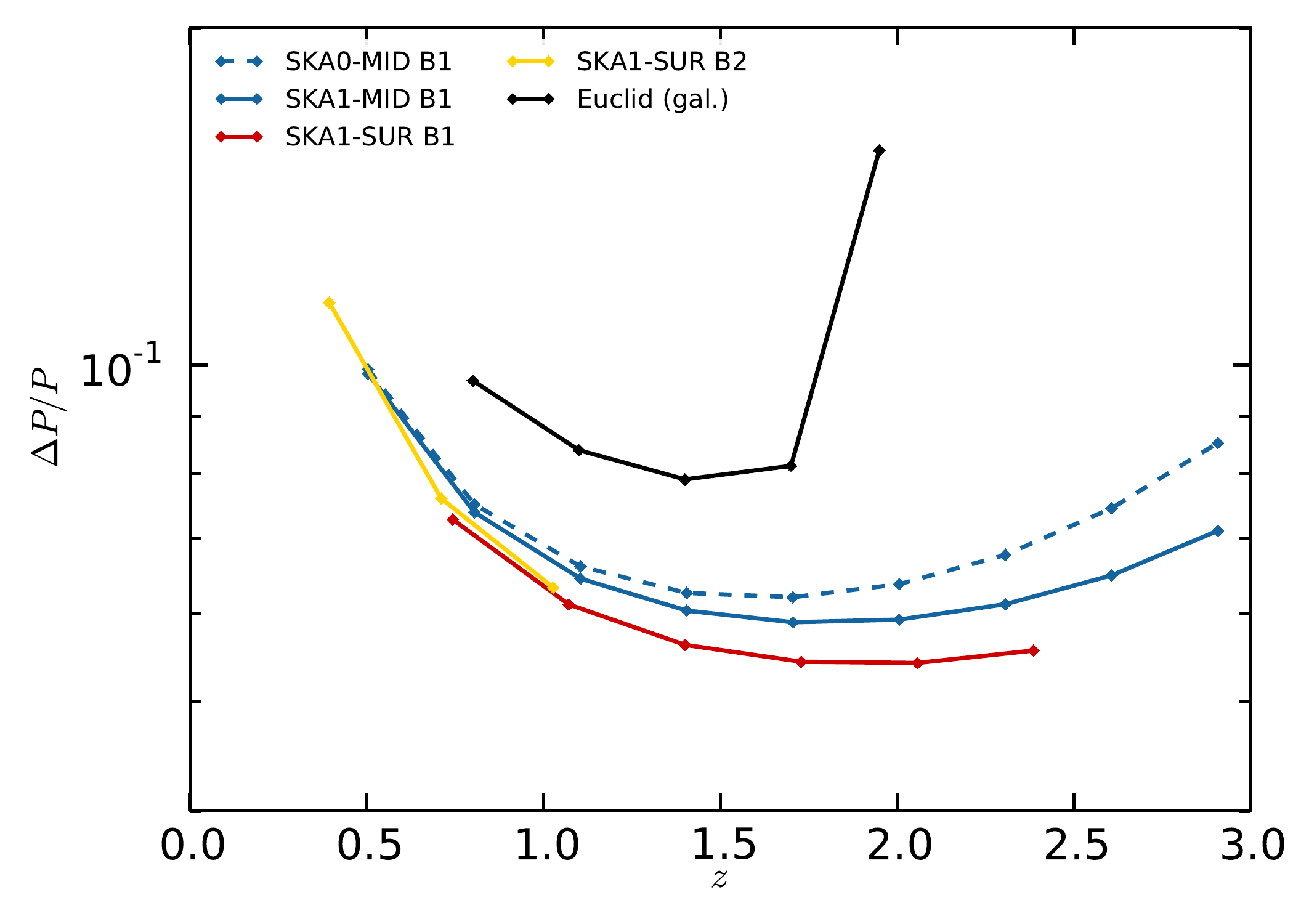}
  \caption[x]{{\bf Left:} Constraints (noise over signal) from SKA HI IM surveys for BAO scales ($k\sim 0.074$ Mpc$^{-1}$) as a function of redshift. Dashed line shows the BAO detection threshold. Assumptions: 10,000 hours observation, 25,000 deg$^2$ survey and bins of $dz=0.1$, except for SKA1-MID in interferometer mode and SKA-LOW where 1,000 deg$^2$ and $dz=0.3$ was taken. The results for SKA0 band 2 (low z), where only 50\% of the dishes are used, was not shown as the results are very similar to SKA1. The lower green curve shows what would be expected from a SKA2 IM survey (in interferometer mode) optimised for high-z. The grey curve shows what can be expected for a two-year H$_\alpha$ galaxy survey with similar depth as Euclid but over a smaller sky area. 
  {\bf Right:} Constraints (noise over signal) from SKA HI IM surveys for large scales, past the equality peak ($k\sim 0.01$ Mpc$^{-1}$) as a function of redshift. A value below $1$ would imply a detection. For SKA1-SUR band 2, the available 500 MHz bandwidth was chosen at the low end of the band in order to probe higher redshifts. SKA1-MID band 2 is not shown as it is constrained to low redshifts ($z<0.5$) with the current band specs. Dashed line indicates what can be achieved with SKA0 (50\% of SKA1) which is quite similar to SKA1. Note that, in order to be as generic as possible, we did not include the foreground contamination in this analysis since the results will depend on the cleaning method adopted. The foreground residuals should degrade these constraints, specially on scales of the order of the frequency band.} \vspace{-2.0em}
\label{fig:dP}
\end{center}
\end{figure*}

While individual galaxies cannot in general be resolved, each telescope pointing measures the integrated emission from many galaxies, making the total signal easier to detect and reducing the shot noise. All that is required is to obtain sufficient flux sensitivity to detect the integrated 21cm emission and to have sufficient resolution to resolve the required scales at a given redshift. Figure \ref{fig:dP} summarises the expected constraints from the SKA HI IM surveys for two relevant target scales: the BAO scale at $k\sim 0.074$ Mpc$^{-1}$ and a very large scale, past the equality peak at $k\sim 0.01$ Mpc$^{-1}$. We see the huge constraining power of these surveys. In particular, due to the large volumes probed, they will be unmatched on ultra-large scales. Even at BAO scales, both SKA1-MID and SUR present constraints not far from Euclid while only using a $\sim 2$ year survey (the full Euclid requires about 5 years). Moreover, SKA1-LOW will be able to make a detection at $z\sim 4$ which again will be an unique feature.
%%%%%%%%%%%%%
%%%%%%%%%%

\begin{figure*}
\begin{center} 
  \includegraphics[width=0.495\textwidth]{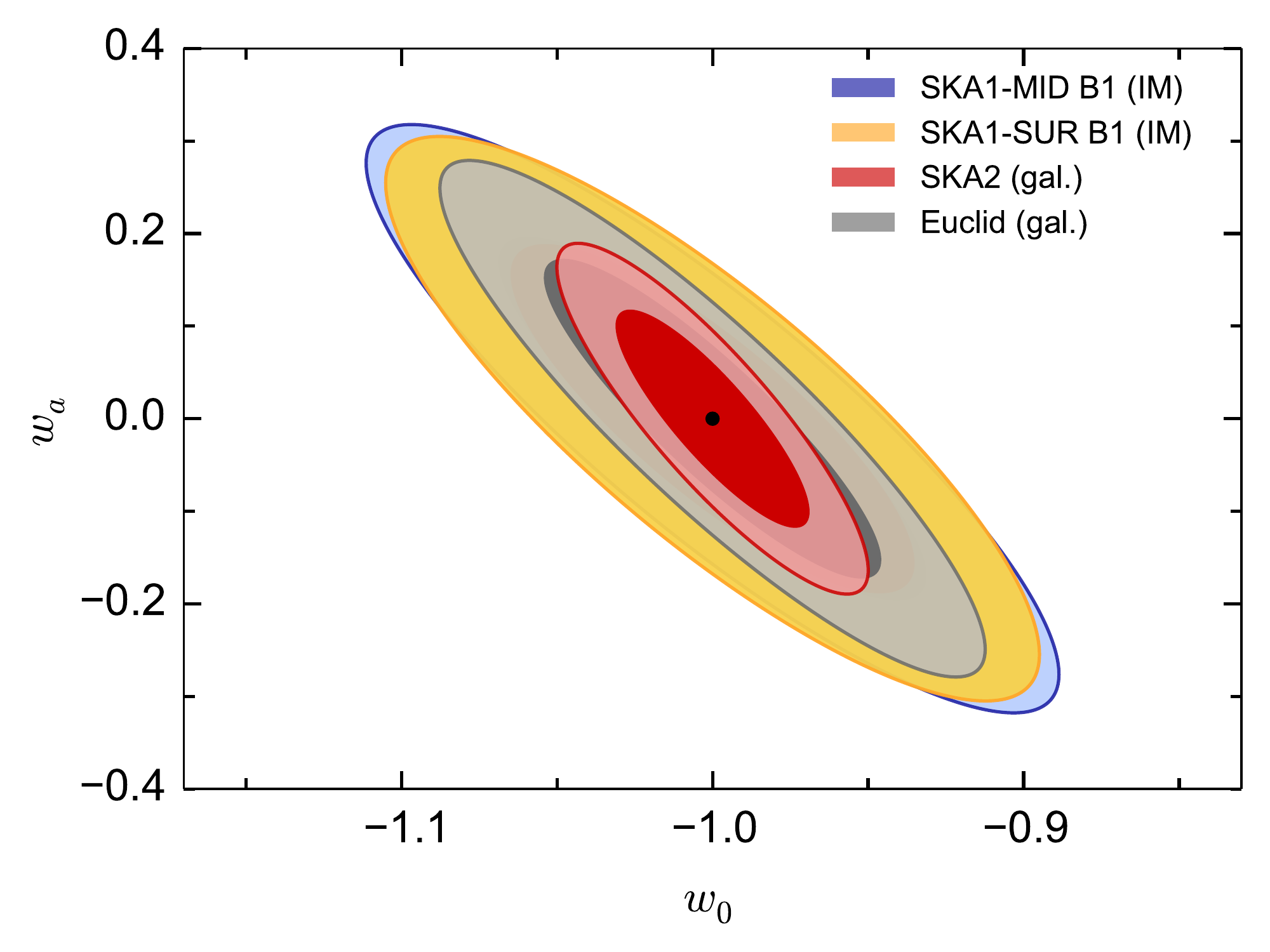}
  \includegraphics[width=0.495\textwidth]{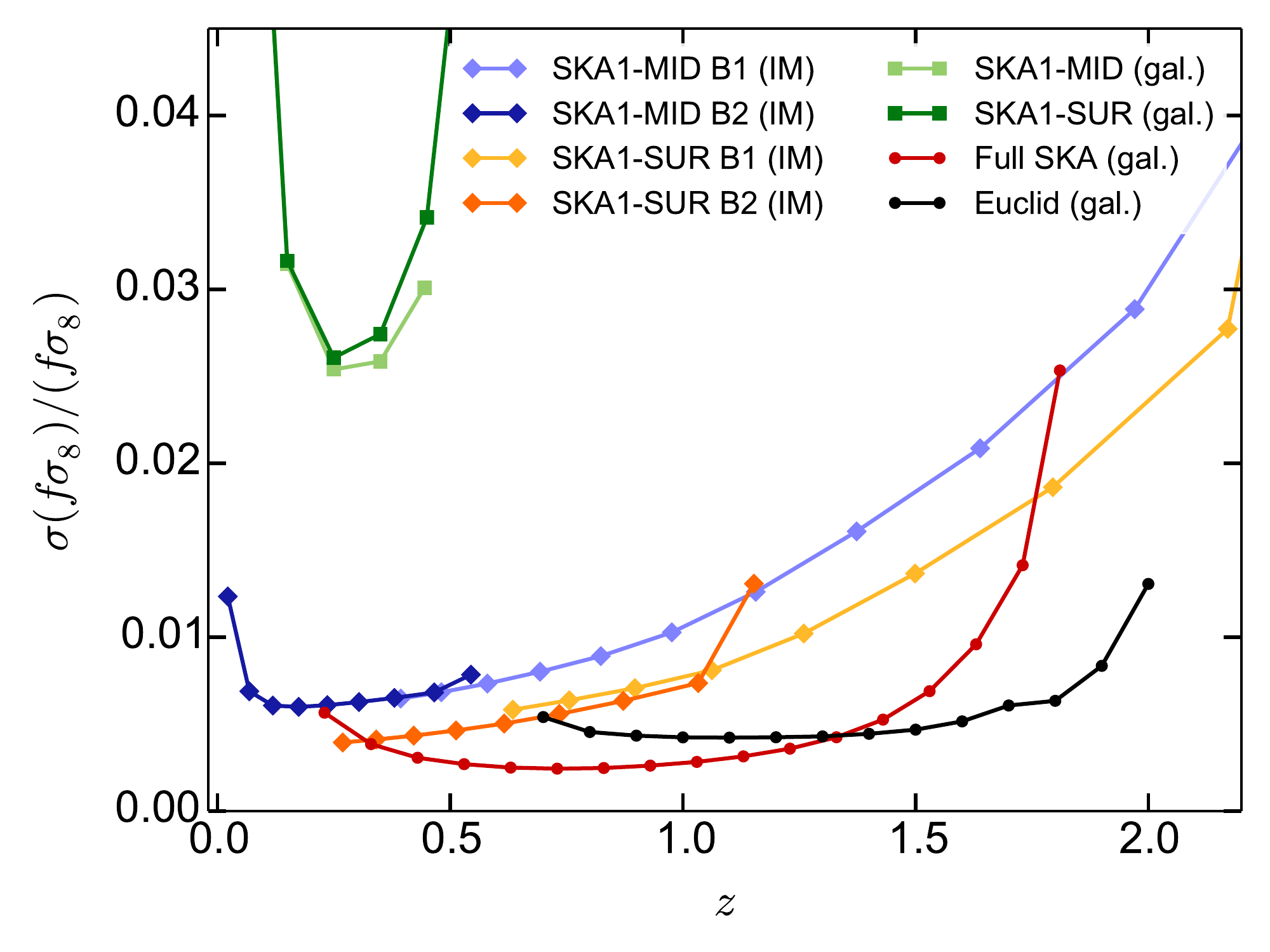}
  \caption{{\bf Left:} Predicted constraints from SKA on dynamical dark energy parameters. We show predicted constraints from SKA1 IM and SKA2 galaxy, compared with predictions for Euclid.  
  {\bf Right:} Predicted constraints from SKA on the unparameterized growth function $f \sigma_8$ from the SKA1 (galaxy and IM) and the SKA2 galaxy survey, compared with predicted constraints coming from the Euclid galaxy survey. Both constraints include Planck+BOSS priors.} \vspace{-1.5em}
\label{fig:waw0}
\end{center}
\end{figure*}

For the BAO scales (see Table \ref{table:rms} and Fig. \ref{fig:dP}, left panel),  the angular resolution of the Phase 1 SKA dishes is such that these scales are best matched to an autocorrelation survey at low redshift, and an interferometric survey at higher redshift. Measurements of the equation of state are most critical at lower redshifts, $z \lesssim 1.5$, where dark energy begins to dominate the cosmic expansion. \citet{bull} show that a 10,000 hour and 25,000 deg$^2$ autocorrelation survey on either SKA1-MID or SUR will be capable of producing high-precision constraints on $w$, bettering all existing surveys due to its large survey area (see Fig. \ref{fig:waw0}).
%%%%%%%%%%%%%
%%%%%%%%%%
While the resulting dark energy `figure of merit' is a factor of $\sim 3$ worse than forecasts for a future Euclid galaxy redshift survey when combined with Planck CMB data and BOSS low-redshift BAO measurements (since Euclid cannot probe redshifts below 0.7), a phase 1 IM survey will nevertheless be of great utility in superseding other low-$z$ measurements in the joint analyses that will produce the best constraints on $w$.

Another important quantity that can be derived from BAO measurements is the spatial curvature, $\Omega_K$, which describes the global geometry of the observable Universe. A key prediction of the prevailing inflationary theory of the early Universe is that the spatial curvature should be extremely small. Current constraints (e.g. \citealt{Ade:2013zuv}) find $|\Omega_K| \lesssim 10^{-2}$, but a precision measurement at the $\sim\mathrm{few}\,\times\, 10^{-4}$ level is needed to really put pressure on inflationary models (e.g. \citealt{Kleban:2012ph}). In combination with Planck CMB data, an SKA IM survey would be able to approach this value, measuring
\be
 |\Omega_K| < 10^{-3}
 \ee
  with 68\% confidence \citep{2014arXiv1405.1452B}.

%%%%%%%%%%%
\subsection{Growth of structure}

Viewed in redshift space, the matter distribution is anisotropic due to the distorting effect of peculiar velocities in the line of sight direction. Coherent peculiar velocities on large scales encode information about the history of the growth of structure in the Universe through their dependence on the linear growth rate, $f(z)$, which can be measured from the degree of anisotropy of the redshift-space correlation function. The growth rate is directly related to the strength of gravity, and so is an extremely useful tool for probing possible deviations from general relativity that have been invoked as an alternative to dark energy to explain cosmic acceleration.

Intensity mapping and galaxy surveys do not measure the linear growth rate directly, but are instead sensitive to simple combinations of $f(z)$, the bias $b(z)$, and the overall normalisation of the power spectrum $\sigma_8(z)$. A reasonable choice of parametrisation is to take the combinations $(f \sigma_8, b \sigma_8)$. As shown in \citet{Raccanelli}, a 10,000 hour and 25,000 deg$^2$ SKA phase 1 intensity mapping autocorrelation survey will be capable of measuring $f \sigma_8$ with high precision over a wide redshift range, obtaining sub-1\% constraints in the range $0.05 \lesssim z \lesssim 1.0$ with Band 2 of SKA1-MID or SUR, and reaching out to $z \approx 2.0$ with $\sim 4\%$ precision using Band 1 of MID/SUR (see Fig. \ref{fig:waw0}).

At low redshifts, these figures are highly complementary to (e.g.) a Euclid galaxy redshift survey, which should obtain $\sim 0.5\%$ measurements of $f \sigma_8$ in the interval $0.7 \lesssim z \lesssim 2.0$. By comparison, SKA1-MID/SUR will have $\sim 0.5\%$ measurements for $z \approx 0.3$ -- $0.7$.

%%%%%%%%%%%%%
\subsection{Probing ultra-large scales}

\begin{figure}
\centering
\vskip -0.8in
\includegraphics[width=0.49\textwidth]{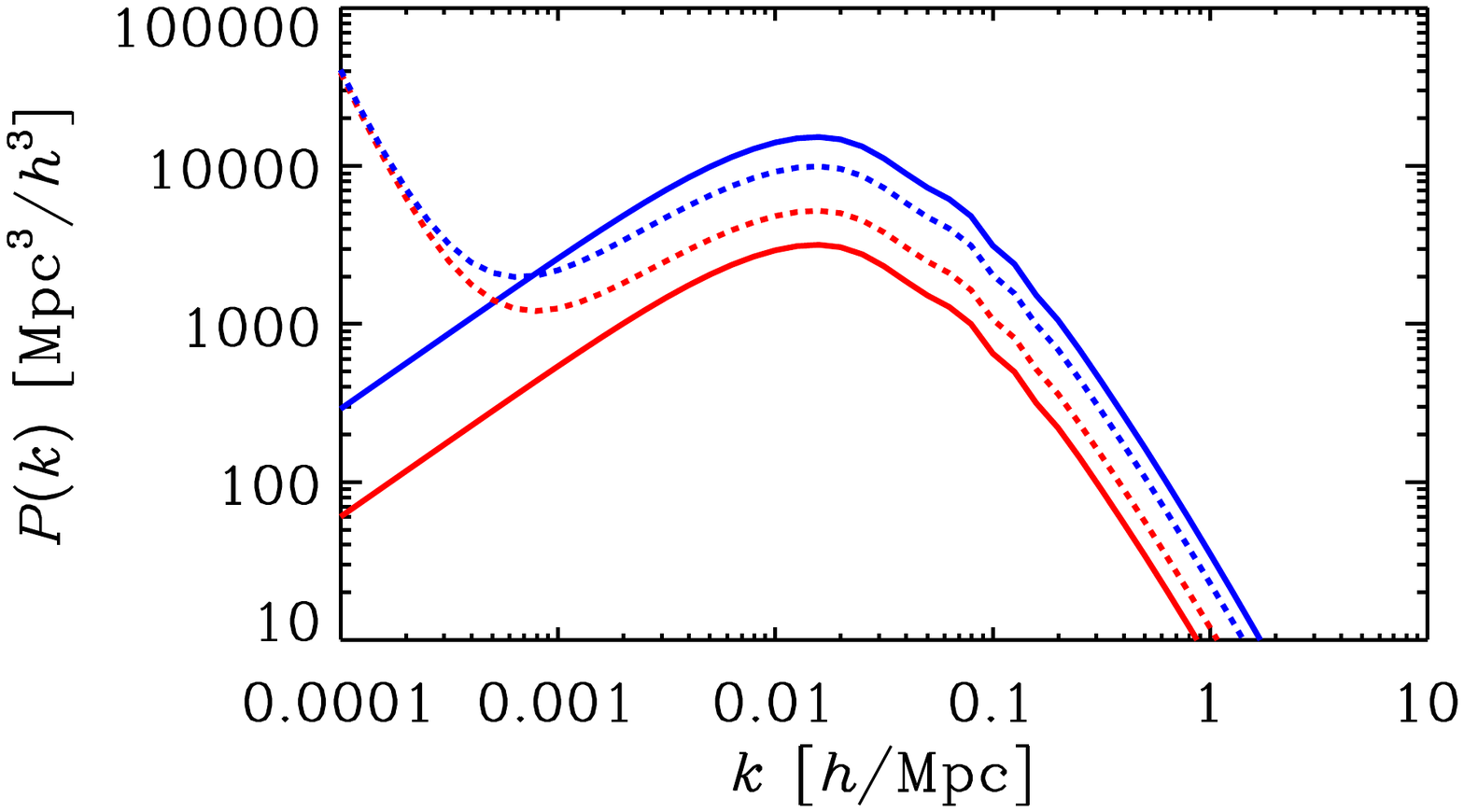}
\includegraphics[width=0.49\textwidth]{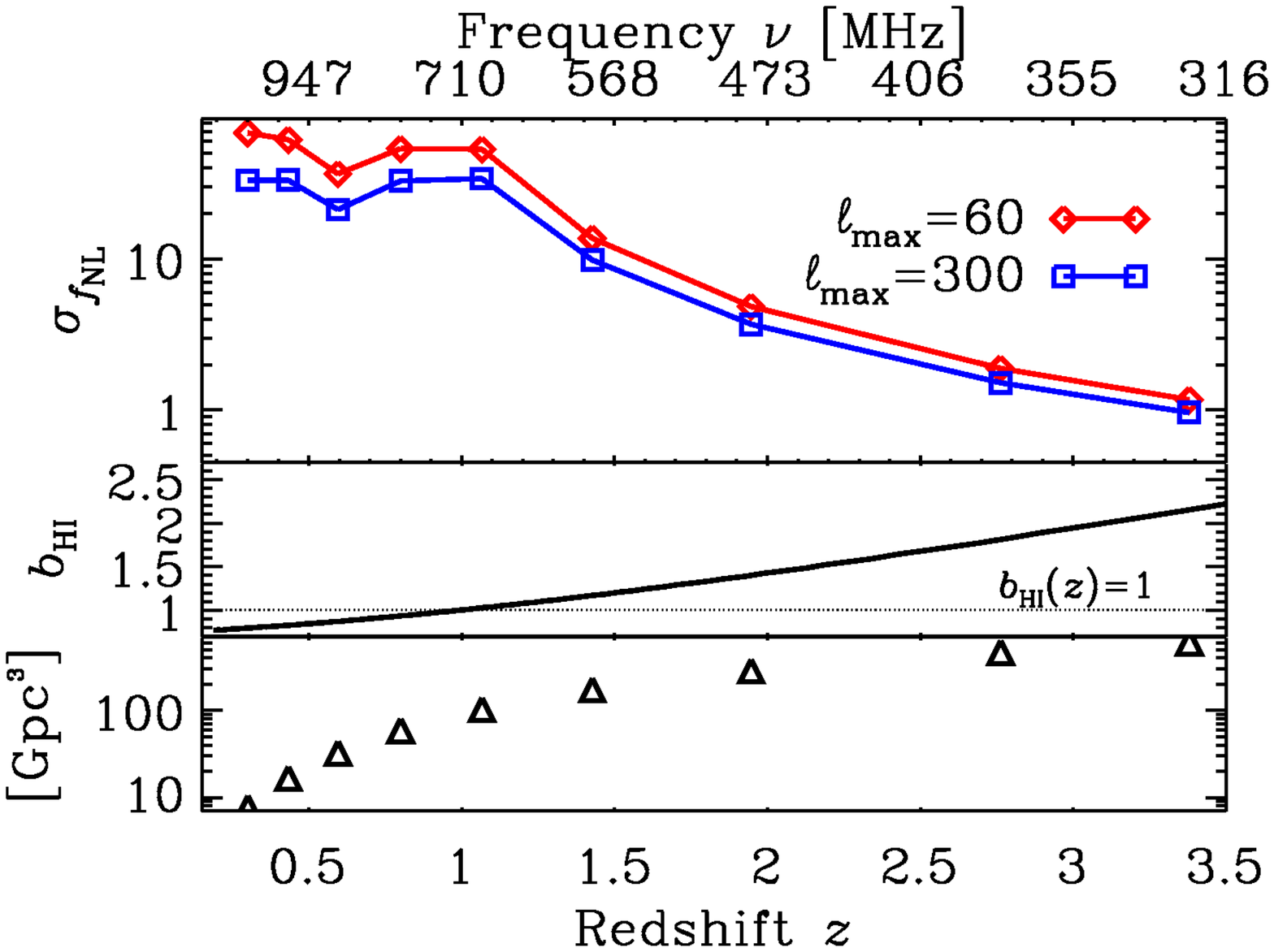}
\vskip -1in
\caption{{\em Left:} Power spectrum of dark matter (solid) and HI (dashed) at $z=0.4$ (blue, top) and $z=2.5$ (red, bottom), with $f_{\rm NL}=10$. {\em Right:}
Forecast $1\sigma$ error on $f_{\rm NL}$ (top); HI Gaussian bias (middle); effective IM survey volume (bottom). From \citet{2013PhRvL.111q1302C}.}\label{nong}
%\vskip -0.2in
\end{figure}

As briefly mentioned above, there is important information that can be extracted from the ultra-large scale modes of order and above the cosmological horizon (see Fig. \ref{fig:dP}, right panel).
We refer the reader to \citet{Camera:AASKA14} and references therein for an extensive description of the ultra-large scale effects briefly mentioned here, as well as to the ways by which the SKA will be able to tackle successfully the technical problems arising when trying to access those scales. 

One of the most important features on horizon scales is primordial non-Gaussianity.
Many models of inflation predict a small amount of non-Gaussianity in the statistical distribution of primordial fluctuations. This produces a signal in the bispectrum, but also in the power spectrum -- since primordial non-Gaussianity induces a scale-dependent correction to the Gaussian bias: $b \to b+\Delta b$. This correction grows on large scales as $\Delta b \propto f_{\rm NL}k^{-2}$ for primordial non-Gaussianity of the local type, where $f_{\rm NL}$ is the non-Gaussian parameter.

In \citet{2013PhRvL.111q1302C}, an analysis is given of the constraining power of IM surveys over non-Gaussianity; their results are summarised in Fig. \ref{nong}. This shows that the forecast errors on $f_{\rm NL}$ can be taken down towards $\sigma_{f_{\rm NL}} \lesssim3$ for a deep enough survey with sufficient dishes.  We recast their analysis according to the updated specifics of Table~\ref{tab:surveys}, and adopt a SKA1-MID IM survey operating for $10,000$ hours at a system temperature of $20$ K. The chosen bandwidth is therefore $350-1050$, where we keep the last, high-frequency bins between $1000$ and $1050$ MHz for foreground removal. The bandwidth is further subdivided into constant frequency bins of $10$ MHz width, collected into `chunks' of 20 by 20 bins in order to construct a 65 by 65 tomographic matrix. (To deal with the large number of bins, we use a block diagonal tomographic matrix where we correct for the overlapping, as described in \citealt{2013PhRvL.111q1302C}.) Such a configuration eventually yields a constraint on the primordial non-Gaussianity parameter 
\begin{equation}
\sigma_{f_{\rm NL}}= 2.3, 
\end{equation}
namely more than three times better than the current constraint from Planck (using the large-scale structure convention).

SKA IM surveys will also allow us to test  Einstein's theory of general relativity for the first time on horizon scales. One of the most interesting effects predicted by general relativity is the correction to the standard Newtonian approximation for the  observed galaxy overdensity. The Kaiser redshift-space distortion term is a relativistic correction that is significant on small scales. Further relativistic corrections include other redshift terms (Doppler and gravitational), Sachs-Wolfe (SW) type terms, and integrated contributions -- from weak lensing magnification, time-delay and ISW terms (see \citet{Camera:AASKA14}): 
\begin{equation}
\delta^{\rm obs}_{{T_b}}=(b_{\rm HI}+\Delta b)\delta-{(1+z)\over H} n^i\partial_i(\mathbf{n}\cdot \mathbf{v})+\delta_{\rm GR},
\end{equation}
The terms in $\delta_{\rm GR}$ grow on ultra-large scales and need to be accounted for.

It turns out that for IM, some of these corrections are strongly suppressed; e.g. the weak lensing magnification and the integrated time-delay terms are zero for IM \citep{Hall:2012wd,2013JCAP...10..015D}. However, the Doppler terms and the ISW contribution remain and need to be incorporated.

At the same time as testing general relativity, one can also test specific modified gravity theories which deviate from general relativity on large scales. SKA IM surveys should enhance considerably the current constraints on parameters that describe modified gravity solutions; this is an area for future work.

%%%%%%%%%%%%%%%%%%%
\subsection{Weak Lensing}

\begin{figure*}
\centering
\includegraphics[trim={2cm 1cm 1.5cm 1.5cm}, clip, width=0.45\textwidth]{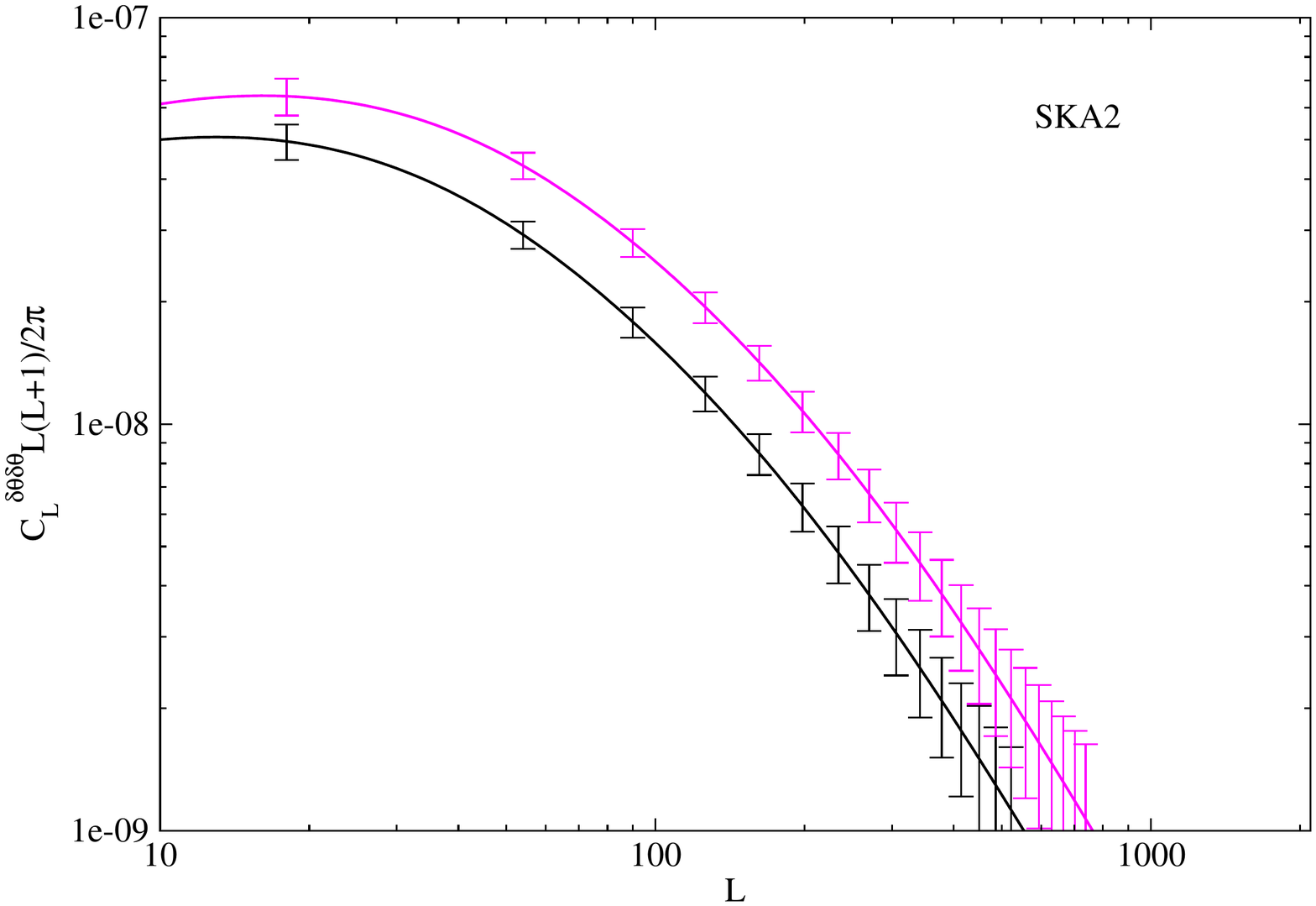}
\includegraphics[trim={2cm 1cm 1.5cm 1.5cm}, clip, width=0.45\textwidth]{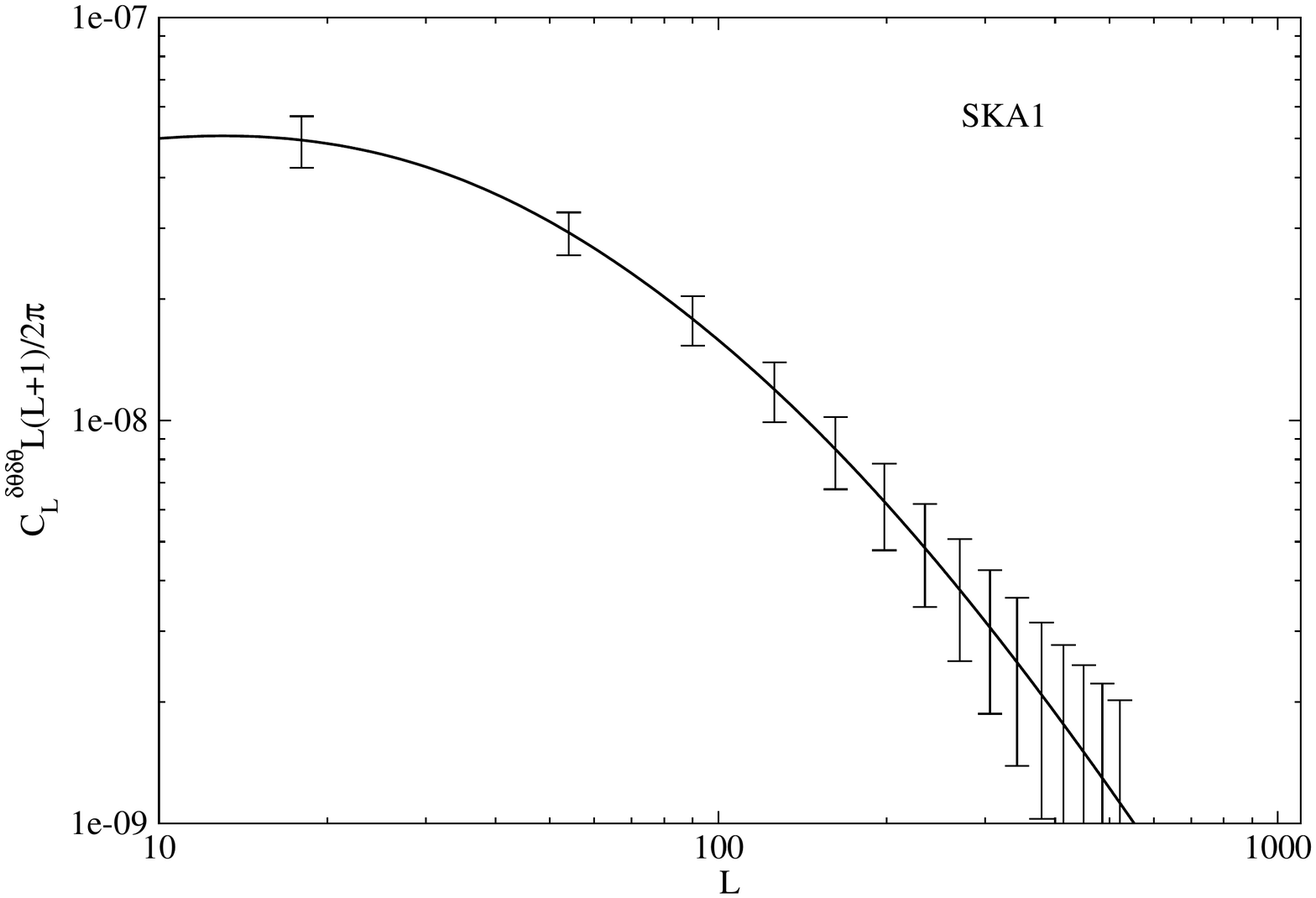}
\caption{{\it Left}: Displacement field power spectrum for $z_s=2$ (solid black line) and $z_s=3$ (solid magenta line) and the corresponding measurement errors using the SKA2 specifications and assuming no redshift evolution in the HI density. {\it Right}: Displacement field power spectrum for $z_s=2$ and the corresponding measurement errors using the SKA1 specifications and an increasing $\Omega_{\rm HI}(z)$ with redshift. }
\label{fig:CL}
\end{figure*}

HI intensity mapping can also be used to measure weak gravitational
lensing.  Gravitational magnification will have an effect on the clustering
properties of galaxies that is coherent over a large range in
redshift.  The effect can be detected by applying a quadratic estimator
to the brightness temperature maps.  HI intensity mapping with SKA would allow
for weak lensing measurements at higher redshifts than are possible
with more traditional weak lensing methods based on the shearing of
galaxy images in the visible.  The technique is more fully
described in the Weak Lensing chapter of this proceeding and in
\citet{Pourtsidou:2014,Pourtsidou:2013hea}.  

The signal-to-noise of such a measurement is strongly dependent on the
HI density at $z \sim 2 - 3$ which is not yet strongly constrained by
observations (see figure \ref{fig:signal}).  For a conservative
  assumption of no evolution in $\Omega_{\rm HI}(z)$, we find that
  SKA-MID phase 2 should be able to measure the shape of lensing power
  spectrum and its evolution between $z = 2$ and 3 (see
  Fig.~\ref{fig:CL},  left panel).  This assumes a
  20,000 sq.deg. survey.   

If  $\Omega_{\rm HI}(z)$ increases by a factor of $\sim 5$ by
redshift 3, as seen in the DLA observations from \citet{Peroux:2001ca},
significantly higher signal-to-noise can be achieved and SKA phase 1
should be capable of measuring the lensing power spectrum and its
evolution with high accuracy, as shown in Fig.~\ref{fig:CL}, right
panel.  This would be a unique  probe of the expansion history of the 
universe and gravity at this range in redshift.

\section{Technical challenges}

The idea of using intensity mapping to reconstruct the large scale structure of the universe bring radio astronomy back to what has been one of its greatest successes -- mapping out cosmological diffuse emission. Indeed, tremendous progress has been made in mapping out the cosmic microwave background (CMB) and the hope is that many of the techniques developed there may inform us on how best to proceed. We will now briefly address some of the problems that need to be tackled if we are to move forward with this technique.

For a start it is clear that different redshift ranges will require different observation "modes": at high redshift it is preferable to use interferometers while at low redshifts it should be more efficient to work with single dishes (or "auto-correlation" mode in the parlance of radio astronomers). This is not a watertight rule. For example, with the clever use of phase arrays or cylindrical arrays, it should be possible to construct interferometers with short baselines and large fields of view hence accessing larger wavelengths at lower redshifts. But, for now, this separation of scales/redshift is useful in guiding us through the issues.

At low redshifts one needs to perform a classic CMB-like observation which is to raster scan the sky building up a rich set of cross-linked scans that cover as much area as possible with as much depth as is necessary. The key problems are then dealing with the long term drifts in the noise (the $1/f$ noise which is ubiquitous in such experiments) and accurately calibrating the overall signal. Usually, the first problem can be dealt with sufficiently fast scan speed (such that the bulk of the signal is concentrated in frequency in the regime where the $1/f$ has died off and the noise is effectively uncorrelated) but this can be difficult to achieve with large dishes such as the ones that are envisaged in current and future experiments. For some setups, the fortuitous configuration of elevation and location mean that drift scanning may lead to a fast enough scan speed. 

With regards to calibrating single dish experiments, this is a source of major concern. Major systematic effects to be tackled are spillover and sidelobe pickup as well as gain drifts. Again, these are issues that have been tackled successfully in the analysis of CMB data although novel approaches can be envisaged. So, for example, the BINGO experiment \citep{2013MNRAS.434.1239B} propose to use a partially illuminated aperture and a fixed single dish, minimising the problems that arise from moving parts. Another intriguing possibility is, for a cluster of single dishes working in autocorrelation mode, to use the cross correlation data for calibrating off known sources. This means that in principle, calibrating the gains should be straightforward using the interferometer data since the high resolution will allow access to a good sky model.

In the case of interferometric measurements, the challenge is to capture as much of the long wavelength modes as possible. The largest wavelength is set by the smallest baseline which implies that arrays with large dishes will not adequately sample BAO scales at low redshifts in interferometer mode. To mitigate this problem, one can work with dense aperture arrays which can be a possible design for SKA2 (or just use smaller dishes and pack them closer together). This results in smaller baselines and a larger field of view for the interferometer, but reduces its total effective collecting area (and thus its sensitivity) if we want to maintain the number of correlations low. Alternatively, reflector designs have been proposed -- for example, long cylindrical reflectors with many closely-spaced receivers installed along the cylinder \citep{Shaw:2014khi}. This provides a large number of short baselines, and a primary beam that is $\sim\!\!180^\circ$ in one direction but much narrower along the orthogonal direction.

%===============================================================================
\section{Conclusions}
%===============================================================================

Neutral hydrogen (HI) intensity mapping is set to become a leading cosmology probe during this decade. Intensity mapping at radio frequencies has a number of advantages over other large-scale structure surveys methodologies. Since we only care about the large-scale characteristics of the HI emission, there is no need to resolve and catalogue individual objects, which makes it much faster to survey large volumes. This also changes the characteristics of the data analysis problem: rather than looking at discrete objects, one is dealing with a continuous field, which opens up the possibility of using alternative analysis methods similar to those applied (extremely successfully) to the CMB. Thanks to the narrow channel bandwidths of modern radio receivers, one automatically measures redshifts with high precision too, bypassing one of the most difficult aspects of performing a galaxy redshift survey.

These advantages, combined with the rapid development of suitable instruments over the coming decade, should turn HI intensity mapping into a highly competitive cosmological probe. One of the key instruments that can be used for this purpose is phase I of the SKA. A large sky survey with this telescope should be able to provide stringent constraints on the nature of dark energy, modified gravity models and the curvature of the Universe. Moreover, it will open up the possibility to probe Baryon Acoustic Oscillations at high redshifts as well as ultra-large scales, beyond the horizon size, which can be used to constrain effects such as primordial non-Gaussianity or potential deviations from large-scale homogeneity and isotropy. 

Several challenges will have to be overcome, however, if we want to use this signal for cosmological purposes. In particular, cleaning of the huge foreground contamination, removal of any systematic effects and calibration of the system. Foreground cleaning methods have already been tested with relative success taking advantage of the foreground smoothness across frequency but novel methods need to be explored in order to deal with more complex foregrounds. Other contaminants, such as some instrumental noise bias that shows up in the auto-correlation signal, can in principle be dealt with the same methods. Ultimately, we should deal with the cleaning of the signal and the map making at the same time. This will require even more sophisticated statistical analysis methods and it will be crucial to take on such an enterprise in the next few years in order to take full advantage of this novel observational window for cosmology.

\vspace{0.5cm}

\noindent{\it Acknowledgments:} --- MGS and RM are supported by the South African SKA Project and the National Research Foundation. PB is supported by European Research Council grant StG2010-257080. RM is supported by the UK Science \& Technology Facilities Council (grant ST/K0090X/1).

%===============================================================================
% BIBLIOGRAPHY
%===============================================================================

\bibliographystyle{apj}
\bibliography{HI_IM}

%\begin{thebibliography}{99}

%\bibitem{Pourtsidou:2014a} 
%Pourtsidou, A., \& Metcalf, R.~B., 2014, MNRAS, 439, L36
%\bibitem{Pourtsidou:2014b} 
%Pourtsidou, A., \& Metcalf, R.~B., 2014, submitted to MNRAS. 
%\bibitem{Peroux:2001ca}
%Peroux, C., McMahon, R.~G., Storrie-Lombardi, L.~J., Irwin, M.~J., 2003, MNRAS, 346, 1103

%\end{thebibliography}

\end{document}